\documentclass[sigconf,screen]{acmart}

\usepackage{tikz}
\usepackage{amsmath}
\usepackage{multirow}
\usepackage{makecell}
\usepackage{enumitem}

\usepackage{booktabs}   
\usepackage{subcaption} 
\newcommand{\asplossubmissionnumber}{1601}

\usepackage[normalem]{ulem}
\usepackage{xspace}
\usepackage{color}
\usepackage{listings}
\usepackage{tikz}
\usepackage{listings}
\usepackage{graphicx}
\usepackage{mathpartir}
\usepackage{float}
\usepackage{textcomp}
\usepackage{pythonhighlight}
\usepackage{algorithm}
\usepackage{tabularx}
\usepackage{algpseudocode}
\usepackage{pervasives}
\usepackage{balance}

\setcopyright{acmcopyright}
\acmPrice{15.00}
\acmDOI{10.1145/3503222.3507778}
\acmYear{2022}
\copyrightyear{2022}
\acmSubmissionID{asplos22main-p1601-p}
\acmISBN{978-1-4503-9205-1/22/02}
\acmConference[ASPLOS '22]{Proceedings of the 27th ACM International Conference on Architectural Support for Programming Languages and Operating Systems}{February 28 -- March 4, 2022}{Lausanne, Switzerland}
\acmBooktitle{Proceedings of the 27th ACM International Conference on Architectural Support for Programming Languages and Operating Systems (ASPLOS '22), February 28 -- March 4, 2022, Lausanne, Switzerland}

\begin{document}

\date{}

\title[Breaking the Computation and Communication Abstraction Barrier in Distributed ML Workloads]{Breaking the Computation and Communication Abstraction Barrier in Distributed Machine Learning Workloads}

\author{Abhinav Jangda}
\affiliation{
    \institution{University of Massachusetts Amherst}
    \country{United States}
}
\author{Jun Huang}
\affiliation{
    \institution{Ohio State University}
    \country{United States}
}
\author{Guodong Liu}
\affiliation{
    \institution{Chinese Academy of Sciences}
    \country{China}
}
\author{Amir Hossein Nodehi Sabet}
\affiliation{%
  \institution{University of California, Riverside}
  \country{United States}
}
\author{Saeed Maleki}
\affiliation{%
  \institution{Microsoft Research}
  \country{United States}
}
\author{Youshan Miao}
\affiliation{%
  \institution{Microsoft Research}
  \country{China}
}
\author{Madanlal Musuvathi}
\affiliation{%
  \institution{Microsoft Research}
  \country{United States}
}
\author{Todd Mytkowicz}
\affiliation{%
  \institution{Microsoft Research}
  \country{United States}
}
\author{Olli Saarikivi}
\affiliation{%
  \institution{Microsoft Research}
  \country{United States}
}

\renewcommand{\shortauthors}{A. Jangda, J. Huang, G. Liu, A. H. N. Sabet, S. Maleki, Y. Miao, M. Musuvathi, T. Mytkowicz, O. Sarikivi}


\begin{abstract}
Recent trends towards large machine learning models require both training and inference tasks to be distributed.
Considering the huge cost of training these models, it is imperative to unlock optimizations in computation and communication to obtain best performance.
However, the current logical separation between computation and communication kernels in machine learning
frameworks misses optimization opportunities across this barrier.
Breaking this abstraction can provide many optimizations to improve the performance 
of distributed workloads.
However, manually applying these optimizations requires modifying the underlying computation and communication libraries for each scenario, which is both time consuming and error-prone.

Therefore, we present \tool{}, which 
contains (i) a domain specific language to express a distributed machine learning program in the form of computation and communication operations, (ii) a set of semantics preserving transformations to optimize the program, and (iii) a compiler to generate jointly optimized communication and computation GPU kernels.
Providing both computation and communication as first class constructs allows users to work on a high-level abstraction and
apply powerful optimizations, such as fusion or overlapping of communication and computation.
\tool 
enabled us to optimize data-, model- and pipeline-parallel workloads in large language models with only a few lines of code. 
Our experiments show that \tool 
significantly outperforms state-of-the-art distributed machine learning implementations.
\end{abstract}



\maketitle



\section{Introduction}
\label{sec:intro}

As the trend towards larger machine-learning models continue, from BERT~\cite{bert} with 340 million parameters, GPT-2~\cite{gpt-2} with 1.5 billion parameters, to GPT-3~\cite{gpt3} with 175 billion parameters, model training and inferencing have to be distributed. Moreover, as the computations become resource hungry, optimizing for even the last percentage can have huge benefits in terms of time, energy, and money savings~\cite{gpt3cost,strubell2019energy}.

In machine learning systems today, computation and communication are treated as independent abstractions implemented in different libraries. For instance, computation libraries, such as cuBLAS~\cite{cublas} and cuDNN~\cite{cudnn}, provide optimized tensor algebra operations, while communication libraries, like NVIDIA Collective Communications Library~\cite{nccl}, provide high-performance implementations of collective communication, such as \allreduce. Machine learning frameworks, such as PyTorch~\cite{pytorch}, call computation and communication kernels from these libraries. Thus, in machine learning applications built atop of such frameworks, the computation and communication operations are invoked separately.

While this separation allows independent optimization of computation and communication kernels, breaking this abstraction boundary can unlock new optimizations that are otherwise not feasible. These optimizations include the following. \emph{Interface} optimization eliminates a mismatch between the caller and the callee of an abstraction. For example, a machine learning model's parameters are stored in non-contiguous buffers, one buffer per layer and hence, need to copy all buffers into a single buffer before calling a collective communication like \allreduce. This copy can be avoided if the communication operation takes a list of arrays as input instead of requiring a single buffer. 
\emph{Fusion} optimization decreases memory bandwidth usage by generating a single kernel to perform multiple communication and computation operations. 
\emph{Reorder} optimization moves the computation before or after the communication, thereby either distributing the computation or enabling new fusion possibilities. 
Finally, \emph{overlapping} optimization orchestrates multiple computation and communication operations in a fine-grained manner to fully utilize both network and compute resources. We elaborate on this possibility below.  

\begin{figure}[t]
	\centering
  \includegraphics[width=\linewidth]{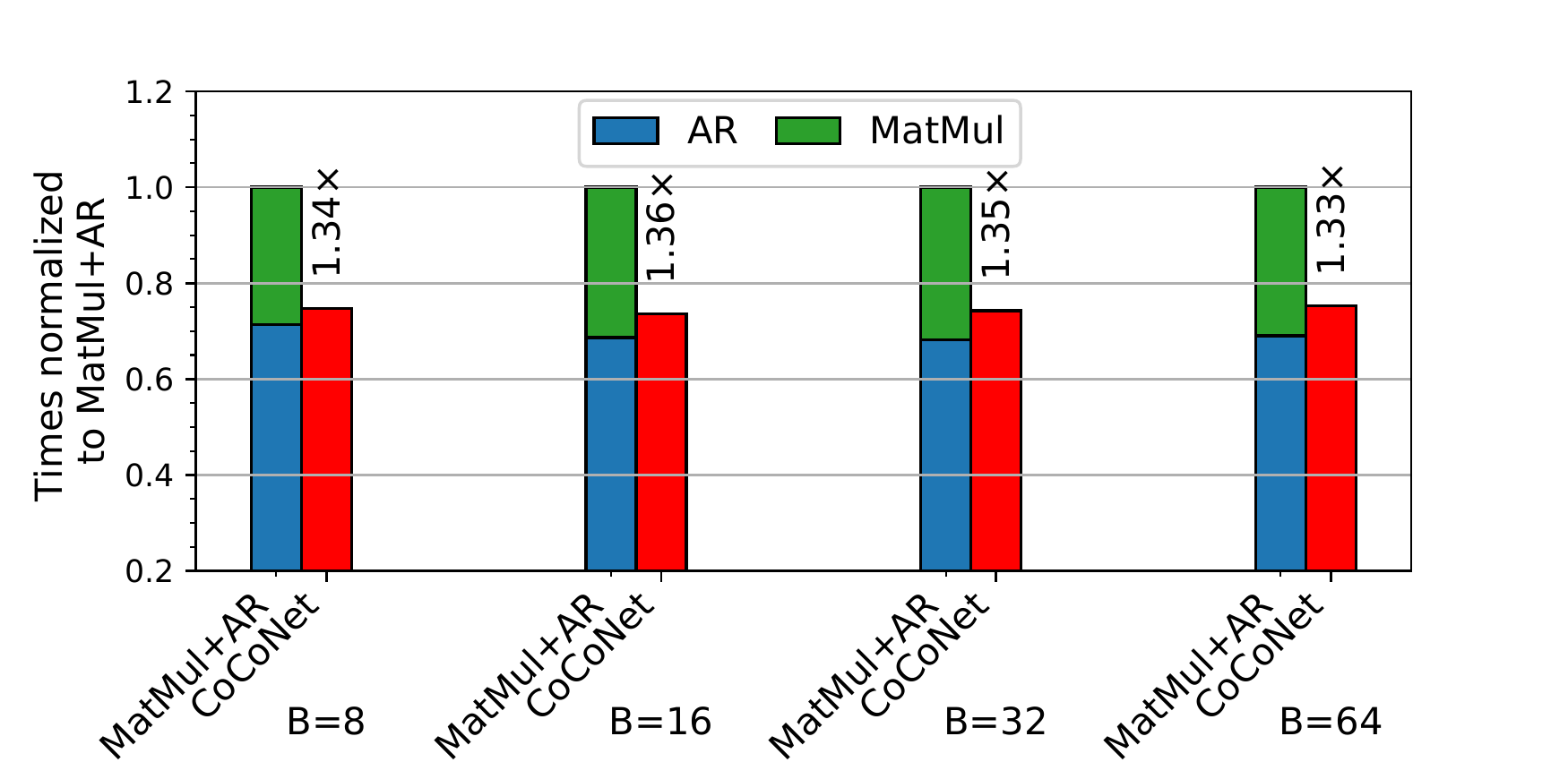}
  \caption{Speedup of co-optimized overlapping over sequential MatMul and \allreduce (for model parallel GPT-2 Model input matrix of [B$\times$1024, 768] and weights of [768, 3072]) on 16 Tesla V100 GPUs.
  \label{fig:matmul-overlap-intro}}
\end{figure}

In model parallelism, which is one of the distributed machine learning approaches, each layer is distributed across multiple GPUs \cite{megatronlm} and the computation for each layer consists of a matrix multiplication (MatMul) on each node followed by an \allreduce. 
The existing implementation of model parallelism calls individually optimized library functions for MatMul and \allreduce.
However, the implementation cannot utilize both network and computation resources simultaneously because the network is idle during MatMul.
We can completely utilize both network and computation resources simultaneously by overlapping the computation of MatMul with the communication of \allreduce in a fine-grained manner.
The idea is to slice the output into smaller chunks and start the \allreduce communication on a chunk as soon as the MatMul kernel has computed it.
To ensure minimum wait time for the \allreduce kernel, we need to schedule the MatMul kernel to compute chunks in the order the \allreduce kernel communicates them.
For instance, in the ring algorithm for \allreduce, the $n^{\text{th}}$ node sends the chunks to the next node in the 
order starting from the $n^{\text{th}}$ chunk. 
As such, the MatMul kernel on the $n^{\text{th}}$ node needs to generate the chunks in this order. 
Furthermore, we need to invoke only one MatMul kernel and \allreduce kernel to avoid the overhead of launching multiple kernels.
Figure~\ref{fig:matmul-overlap-intro} shows that this fine-grained overlapping of MatMul with \allreduce can hide 80\% of the execution time of MatMul and provides 1.36$\times$ speedup.

However, manually writing these optimizations for each scenario is unproductive, for example, the implementation of above overlapping optimization contains $\approx$2k lines of CUDA code.
Thus, in this paper, we show that by carefully designing a \emph{language} for expressing combinations of computation and communication the benefits of existing machine learning framework's abstraction can be maintained while simultaneously allowing a \emph{compiler} to apply powerful optimizations.
\begin{figure*}[t]
	\centering
    \includegraphics[width=\linewidth]{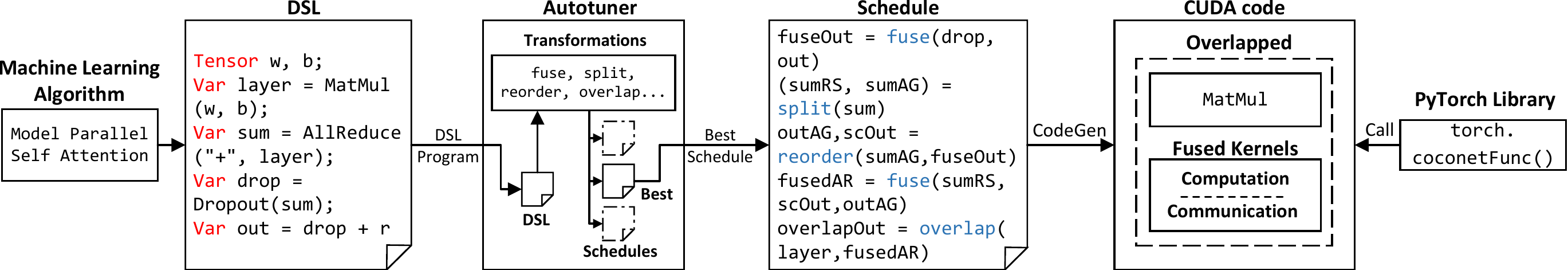}
    \caption{Overview of \tool's workflow. 
    First, a user expresses a machine learning algorithm in the DSL that contains both computation (MatMul) and communication (\allreduce).
    Then, the autotuner applies transformations to optimize the program while keeping the algorithm unchanged, such as fusing \allreduce and Dropout into FusedAllReduce and overlapping this with MatMul.
    Finally, \tool generates custom communication and computation code, which is available through PyTorch.}
    \label{fig:overview}
\end{figure*}
To this effect, we propose \tool\footnote{\tool stands for "{\bf \uline{Co}}mmunication and {\bf \uline{Co}}mputation optimization for neural {\bf \uline{Net}}works.}
for generating co-optimized custom computation and communication kernels.
Figure~\ref{fig:overview} presents the overview of \tool.
\tool includes a domain specific language (DSL) to express programs containing both computation and communication operations.
Inspired by Halide~\cite{halide}, \tool includes a \emph{scheduling} language 
to specify an execution schedule of the program using a set of transformations. 
\tool's \emph{autotuner} automatically applies these transformations to optimize a program by breaking the communication and computation boundary. 
Hence, \tool enables users to quickly generate optimized implementations for 
specific hardware, topology, and data sizes.
\tool's \emph{code generator} automatically generates high-performance computation and communication kernels from a program and its schedule.
We used \tool to optimize data-parallel training, model-parallel inference, and pipeline-parallel inference. 
\tool generated kernels for the Adam~\cite{adam} and LAMB~\cite{lamb} optimizers speeds up
the training time of BERT models by upto 1.68$\times$ and can train BERT 3.9 Billion parameter models using only data parallelism, which is not possible with state of the arts.
\tool's kernels for model parallelism speeds up the inference in BERT 3.9 Billion and GPT-2 8.2 Billion parameter models by upto 1.51$\times$.
\tool's optimized pipeline parallelism kernels speeds up inference times in GPT-2 8.2 Billion and GPT-3 175 Billion parameter models by upto 1.77$\times$.
Our implementation of \tool is available at \url{https://github.com/parasailteam/coconet}.
\section{The \tool DSL}
\label{sec:dsl}
The \tool DSL extends the data representation in existing machine learning frameworks and provides constructs to express both computation and communication. 
The \tool DSL is embedded in C++.
Unifying the expression of computation and communication for distributed machine learning in the same DSL is the foundation to enable optimizations across computation and communication.

In this paper, we follow the MPI~\cite{mpi} terminology: \texttt{RANK} is the process ID of a distributed process, \texttt{GROUP} is a set of concurrent distributed processes, and \WORLD is the \texttt{GROUP} that includes all processes.
\tool supports dividing consecutive ranks into one or more process groups.

\begin{figure}[t]
  \footnotesize
  \begin{lstlisting}[language=DSL]
Tensor w(FP16, [H,H], Sliced(0), WORLD, RANK); |\label{line:mp:inputensor-begin}||\label{line:mp:continuous-tensor}|
Tensor b(FP16, [H], Replicated, WORLD); |\label{line:mp:inputensor-end}| 
Tensor in(FP16, [B,S,H], Sliced(2), WORLD, RANK);
Tensor r(FP16, [B,S,H], Replicated, WORLD);

// layer(FP16, [B,S,H], Local, WORLD, RANK)
Var layer = MatMul(in, w); |\label{line:mp:matmul}|
// sum(FP16, [B,S,H], Replicated, WORLD)
Var sum = AllReduce("+", layer); |\label{line:mp:allreduce}|
// dropout(FP16, [B,S,H], Replicated, WORLD)
Var dropout = Dropout(sum + b, 0.1); |\label{line:mp:pointwise-start}|
// out(FP16, [B,S,H], Replicated, WORLD)
Var out = dropout + r;|\label{line:mp:pointwise-end}|

Execute self_attention({w,in,b,r}, {out});
\end{lstlisting}
	\caption{An example program in \tool. \newline
		(\texttt{B}: batch size, \texttt{S}: sequence length, \texttt{H}: hidden dimension size)
	}
  \vspace{-2em}
    \label{fig:traditional-mp}
\end{figure}

\subsection{Tensor Layout}
\tool extends the concept of a tensor in machine learning frameworks from a single device data into distributed forms.
Besides item datatype, like \texttt{FP32} and \texttt{FP16}, and shape,
a \tool tensor also includes a \emph{layout} that describes the distributed allocation of tensor's data across a set of ranks. 
There are three layouts for a tensor: \emph{sliced}, \emph{replicated}, and \emph{local}.
A \emph{sliced} tensor is equally distributed among all nodes in a group along a specified dimension with \texttt{RANK} identifying the slice for that process.
For example, in Figure~\ref{fig:traditional-mp}, which describes the Megatron-LM~\cite{megatronlm} model parallel logic of Self-Attention layer in \tool, \texttt{w} is sliced among all ranks in \WORLD in the 
first dimension and \texttt{in} is sliced in the third dimension.
A tensor can also be \emph{replicated} across all ranks in a group where it has the same value on each rank and it does not have a rank identifier.
For example, the bias \texttt{b} and the residual connection \texttt{r} are replicated as shown in Figure~\ref{fig:traditional-mp}. 
A \emph{local} tensor has same shape on all ranks but different
values on all ranks. A local tensor requires \texttt{RANK} to 
identify the values. For example, in Figure~\ref{fig:traditional-mp}, \texttt{layer}
is a local tensor that represents the result of MatMul operation.
A \texttt{Scalar} is a zero-dimensional tensor that represents a variable available on all ranks.
We discuss the layout of 
intermediate tensors in the next section.

\subsection{\tool's Operations}
A \tool program inherits the concept of data-flow graph (DFG) from existing machine learning frameworks with operations as vertices and data dependencies as edges.
Operations in \tool can be classified as (i) local computations, such as pointwise computations, matrix multiplication, and convolution, and (ii) cross rank communication operations, such as \allreduce, \allgather, and P2P \send-\recv.
Table~\ref{tab:operations} shows all operations supported by \tool{}.

A \texttt{Var} represents the intermediate tensor obtained after performing an operation. 
In the example of Figure~\ref{fig:traditional-mp}, the linear layer's weight (\texttt{w}) and the input
(\texttt{in}) are sliced across all ranks while the bias (\texttt{b}) and residual
(\texttt{r}) are replicated on all ranks. A \texttt{Var}'s shape and distribution layout are 
inferred based on the operation and inputs to the operation. For example,
line~\ref{line:mp:matmul} performs a MatMul operation on the input (\texttt{in}) and weights (\texttt{w}).
Since MatMul between two sliced tensors produces a local tensor, \texttt{layer} represents the partial result with \textit{local} layout.
At line~\ref{line:mp:allreduce}, \allreduce computes the sum of \texttt{layer} of all ranks and returns a \textit{replicated} tensor with the same values on each rank.
The computations at lines~\ref{line:mp:pointwise-start}--\ref{line:mp:pointwise-end} add the bias, use dropout as an activation, and add the residual. 
At line~\ref{line:mp:pointwise-start}, the addition of \texttt{sum} and \texttt{b} follows
PyTorch's broadcast semantics\footnote{https://pytorch.org/docs/stable/notes/broadcasting.html} by replicating \texttt{b} in all dimensions of \texttt{sum}. 
Thus, the shape and layout of output of these operations are same as \texttt{sum}.
Finally, \texttt{Execute} defines the name, inputs, and outputs of the program. 

\begin{table}
  \small
  \caption{Operations supported by \tool includes all common communication and computation operations. \label{tab:operations}}
  \begin{tabular}{|c|l|}
    \hline
    \textbf{Communication} & AllReduce, AllGather, ReduceScatter, \\
    \textbf{Operations}    & Reduce, Broadcast, P2P Send-Recv \\\hline
    \textbf{Layers} & Matrix Multiplication, Convolution\\ \hline
    \textbf{Activations} & Dropout, tanh, ReLU \\ \hline
    \textbf{Tensor} & $+$, $-$, $*$, $\div$, Norm, ReduceTensor,\\
    \textbf{Operations} & Sqrt, Pow, Update\\\hline 
  \end{tabular}
  \par \bigskip
\end{table}


\subsection{Fused Collective Communication Operations}
\label{sec:fuse-comm-coll}
\tool enables efficient computations on the output of communication by providing fused collective communication operations, such as FusedAllReduce.
Consider the \allreduce in Figure~\ref{fig:traditional-mp} followed by a Dropout (lines ~\ref{line:mp:allreduce}--\ref{line:mp:pointwise-start}).
The abstraction in existing machine learning frameworks requires the output of \allreduce 
to be stored in memory and then re-loaded by Dropout.
FusedAllReduce avoids such stores and loads by directly passing the output of communication to following computations through registers.
In addition to the argument of \allreduce, a FusedAllReduce takes computations as extra arguments.
Section~\ref{sec:code-gen-fused} discusses the implementation of Fused Collective Communication Operations.

\subsection{Overlapping Operations}
\label{sec:overlap-comm-coll}
\tool supports overlapping multiple dependent computation and communication operations using the \texttt{Overlap} construct.
For example, consecutive MatMul and \allreduce in Figure~\ref{fig:traditional-mp} 
(lines~\ref{line:mp:matmul}--\ref{line:mp:allreduce}) can be overlapped to fully utilize both network and computation resources.
Section~\ref{sec:overlap-impl} discusses the implementation of this construct.

\subsection{Custom Operations}
In \tool, the implementation of an operator needs to define three key properties of the operator: (i) syntax, (ii) semantics, and (iii) code generation.
The syntax of an operator is defined using C++ constructors and the semantics are defined by implementing rules to describe the layout and size of the output tensor based on the input tensors.
Finally, the code generation requires implementing a function to generate a call to existing libraries or generate fused GPU kernels.
The implementation of syntax and semantics can be achieved in a few lines of code, however, implementing the code generation for complex operations like Matrix Multiplication and Convolution can potentially take hundreds of lines of code.
Fortunately, in practice the code generation for complex operations can call an optimized implementation of existing libraries.

\begin{figure*}[t]
	\centering
  \includegraphics[width=\linewidth]{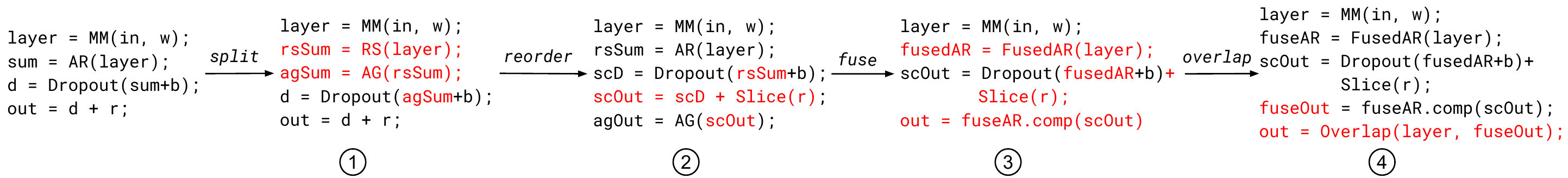}
  \caption{\tool programs produced by performing transformations on the program of Figure~\ref{fig:traditional-mp}. Each schedule can be represented as a standalone program. Lines in \textcolor{red}{red} highlights changes at a step. (\texttt{AR}: AllReduce, \texttt{AG}: AllGather, and \texttt{RS}: ReduceScatter)
  	\label{fig:mp-schedules}}
    \vspace{-1em}
\end{figure*}

\section{\tool Transformations}
\label{sec:schedule}
\tool provides four semantics preserving \emph{transformations} to optimize a program written in the DSL.
All transformations are valid based on rules described in the sections below. 
\tool automatically checks the validity of each transformation based on these rules and throws an error for  an invalid transformation.

We call an order of transformations a \emph{schedule}.
A user can manually specify the schedule to optimize the program.
Additionally, a user can invoke the autotuner to automatically find the best performing schedule for the given problem sizes and the underlying architecture.
Below we present each transformation by applying them on the program from Figure~\ref{fig:traditional-mp} and show equivalent \tool{} programs generated after applying each transformation in Figure~\ref{fig:mp-schedules}.

\subsection{Splitting Communication}
The \texttt{split} transformation breaks a collective communication operation into two communication operations.
One of the two split policies supported by \tool is

\textbf{AllReduce Split RS-AG} splits an \allreduce into a \reducescatter to produce a \sliced tensor and an \allgather on the \sliced tensor to return a \replicated tensor.


\spara{Running Example} The \allreduce in Figure~\ref{fig:traditional-mp} is split into \texttt{rsSum} that does a \reducescatter on \texttt{layer} and \texttt{agSum} that does an \allgather on \texttt{rsSum}.
{
\small
\begin{lstlisting}[language=DSL,numbers=none]
(rsSum, agSum) = split(layer, ARSplitRSAG);
\end{lstlisting}
}

The program \circled{1} of Figure~\ref{fig:mp-schedules} is the implementation of this schedule where the input to Dropout is replaced by \texttt{agSum}.

\spara{\textit{Validity}} Since an \allreduce can always be split to a \reducescatter and an \allgather, this transformation is always valid.

\begin{figure}[t]
	\centering
	\includegraphics[width=.9\linewidth]{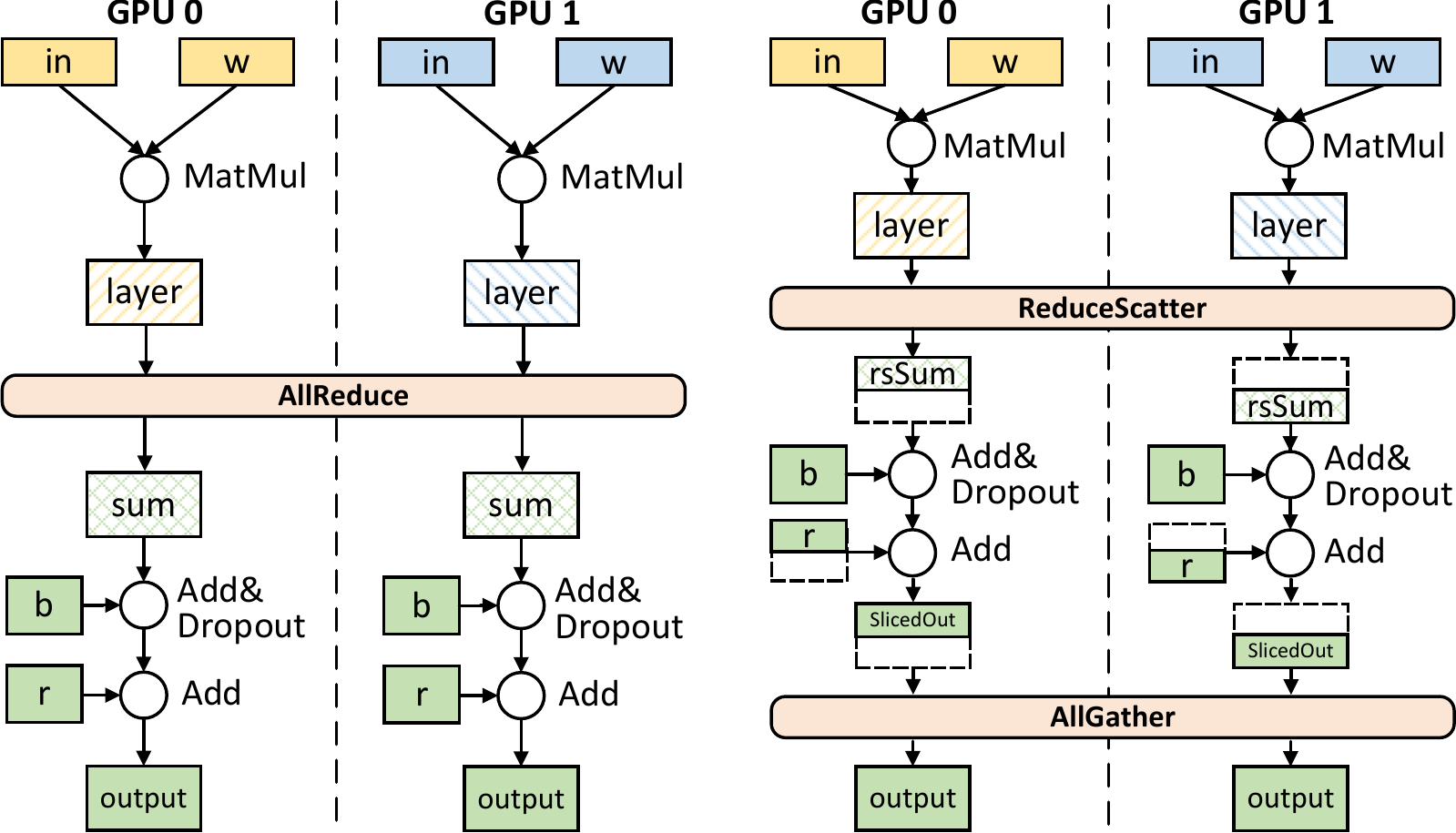}
	\caption{Equivalent programs (from Figure~\ref{fig:traditional-mp}) using AllReduce (on left) or using ReduceScatter + AllGather (on right).}
	\label{fig:model-parallel-using-reducescatter}
\end{figure}

\subsection{Reordering Operations}
The \texttt{reorder} transformation swaps operations with an \allgather or a \broadcast in the DFG of a program.
We explain this transformation for \allgather below: 

\textbf{AllGather Reorder} reorders an \allgather with communication and computation operations. 
This transformation changes the layout of the operations, the input and output of operations, and the input and output of the \allgather.
We explain this transformation below using the running example.


\spara{Running Example}
In Figure~\ref{fig:mp-schedules}, applying the \texttt{reorder} transformation changes the program \circled{1} to \circled{2} by reordering \allgather (\texttt{agSum}) with computations \texttt{d} and \texttt{out}.
The reorder transformation replaces these operations in the DFG with three new operations: \texttt{scD} and \texttt{scOut},
both of which performs sliced computations, and \texttt{agOut}, which gathers the final result of computations.
{
  \small
\begin{lstlisting}[language=DSL,numbers=none]
(scD, scOut, agOut) = reorder(d, out, agSum);
\end{lstlisting}
}
The new sliced computations perform the same operations as original computations with two differences: (i) the output of \allgather used in the computation is replaced by the input of \allgather, and (ii) since the input of \allgather is sliced, 
all tensors input to the computations are also sliced along the same dimension as the input of \allgather.
After reorder, \texttt{scD} performs the same computation as \texttt{d} but \texttt{scD} takes \texttt{rsSum} and \texttt{Slice(r)} as input.
Therefore, the layout of \texttt{scOut} is also sliced while the computation is same as \texttt{out}.
Furthermore, the new \allgather is performed on the outputs of the computations, for example, 
after reorder, the \allgather (\texttt{agOut}) is performed on \texttt{scOut}.
Figure~\ref{fig:model-parallel-using-reducescatter} shows the workflow of this schedule.

\spara{\textit{Validity}} The \texttt{reorder} transformation is valid only if operations being reordered with an \allgather can be sliced along the dimension the \allgather is performed.
The rules of slicing an operation depend on the type of operation and the dimensions of inputs to the operations.
For example, \texttt{d} and \texttt{out} can be sliced because the computations have the same dimensions as \texttt{agOut}.
Section~\ref{sec:opt-workloads} shows how P2P Send can be reordered with an \allgather.

\subsection{Fusing Operations}
\label{sec:sched:fusion}

Fusing multiple computations is a common technique used by existing compilers~\cite{tvm18,distributed-halide,fireiron,polymage-gpu,halide}.
\tool extends this concept to fuse multiple computations and communications in a single operation and provides this capability using the \texttt{fuse} transformation.
Below we explain two fuse policies supported by \tool:

\textbf{Computation Fuse} fuses a series of computations in a single operation that performs all these operations.

\textbf{AllReduce Fuse} fuses a series of \reducescatter, sliced computations, and \allgather operations in a single Fused\allreduce that performs all these operations.

\spara{Running Example}
We can fuse \reducescatter (\texttt{rsSum}), computations (\texttt{scD} and \texttt{scOut}), and \allgather (\texttt{agOut}) in program \circled{2} of Figure~\ref{fig:mp-schedules} into a Fused\allreduce to obtain program \circled{3}.
{
\small
\begin{lstlisting}[language=DSL,numbers=none]
fuseAR = fuse(rsSum, scOut, agOut, ARFuse);
\end{lstlisting}
}
The \texttt{comp} method of \texttt{fusedAR} specifies the computation to be fused with Fused\allreduce and returned \texttt{out} is the output.

\spara{\textit{Validity}} Fusing multiple operations into one operation is valid only if the dependencies in the DFG after fusion are preserved.

\subsection{Overlapping Operations}
\tool provides the \texttt{overlap} transformation to overlap a series of producer-consumer operations to utilize multiple resources of hardware simultaneously.

\spara{Running Example} 
In the program \circled{3} of Figure~\ref{fig:mp-schedules} we overlap the matrix multiplication (\texttt{layer}) with Fused\allreduce (\texttt{fuseAR}) to obtain program in \circled{5}.
{
\small
\begin{lstlisting}[language=DSL,numbers=none]
layerWithAR = overlap(layer, fusedAR);
\end{lstlisting}
}

\spara{\textit{Validity}} Overlapping multiple operations is valid only when all operations have a producer-consumer relationship between them.


\subsection{Automatic Exploration of Schedules}
\tool provides an \emph{autotuner} to automatically explore the space of all schedules of a program and return the schedule that provides the best performance for the underlying architecture and input sizes.
First, the autotuner fuses all pointwise computations up to a pre-defined threshold to decrease the search space and then exhaustively explores the schedule space in a breadth first search manner.
Finally, the autotuner generates code for all schedules in its search space, executes all programs, and returns the schedule with minimum execution time.
Table~\ref{tab:loc-autotuner-time} shows that the autotuner takes only a few seconds to explore the schedule space for all workloads.

\begin{figure}
  \small
  \centering
  \begin{subfigure}[t]{\columnwidth}
\begin{lstlisting}[language=DSL, numbers=left]
Var avg = AllReduce("+", g); |\label{line:adam:avg}|
Var m_ = Update(m, (m*beta1+(1-beta1)*avg));|\label{line:adam:pointwise-begin}||\label{line:update:m}|
Var v_ = Update(v, (v*beta2+(1-beta1)*avg*avg));|\label{line:update:v}|
Var m1 = m_/(1-Pow(beta1, t));
Var v1 = v_/(1-Pow(beta2, t));
Var p_ = Update(p, (p - lr * m1/(Sqrt(v1))));|\label{line:adam:pointwise-end}|

Execute adam({g,p,v,m,lr}, {p_});
\end{lstlisting}
\caption{Traditional implementation where 
               tensors \texttt{g} is local to each rank and \texttt{p},\texttt{m}, and \texttt{v} are replicated on all ranks.}
\label{fig:traditional-adam}
\end{subfigure}
\par\bigskip 
\begin{subfigure}[b]{\columnwidth}
\begin{lstlisting}[language=DSL, numbers=left]
comps = fuse(m_, v_, m1, v1, p_, 
             ComputationFuse);|\label{line:adam-schedule:fuse-comp}|
(rsG, agG) = split(avg, ARSplitRSAG); |\label{line:adam-schedule:split}|
(scComp, agP, agM, agV) = reorder(agG, comps, 
                                  AGReorder);|\label{line:adam-schedule:reorder}|  
asSlice(m); asSlice(v); dead(agM); dead(agV); |\label{line:adam-schedule:slice-m-v}| |\label{line:adam-schedule:remove-m-v-allgather}|
fuseAR = fuse(rsG, scComp, agP, AllReduceFuse);|\label{line:adam-schedule:fuse-allreduce}|
\end{lstlisting}
\caption{An Optimized Schedule. Tensors \texttt{g} is local, \texttt{p} is replicated on all ranks, while 
\texttt{m} and \texttt{v} are sliced among all ranks.}
\label{fig:adam-schedule}
\end{subfigure}

  \caption{Optimizing parameter update using Adam in \tool. 
  The implementation takes four input tensors: parameters (\texttt{p}), gradients (\texttt{g}),
  momentum (\texttt{m}), and velocity (\texttt{v}).
  \label{fig:optimizing-adam}}
  \end{figure}

\section{Distributed Workloads in \tool}
\label{sec:opt-workloads}
We additionally optimized two distributed machine learning workloads using \tool: (i) parameter update using Adam~\cite{adam}, and (ii) point-to-point communication in pipeline parallelism.

\spara{Adam in Data Parallel Training}: Figure~\ref{fig:traditional-adam} shows the traditional implementation of parameter update using Adam.
First, all ranks average the gradients using \allreduce and then perform computations to update the optimizer state and model parameters.
\texttt{Update} updates the values of a tensor and reflects the new values in that position in the DFG (lines~\ref{line:update:m}--\ref{line:update:v}).
Figure~\ref{fig:adam-schedule} presents a schedule that optimizes this by distributing the computation on all ranks in a single kernel.
Line~\ref{line:adam-schedule:fuse-comp} fuses all computations in \texttt{comps}.
Line~\ref{line:adam-schedule:split} splits the \allreduce into a \reducescatter and an \allgather, such that computations take output of \allgather (\texttt{agG}) as input.
Line~\ref{line:adam-schedule:reorder} reorders \allgather with computations, such that,
each rank performs computations on a slice of tensors.
Line~\ref{line:adam-schedule:slice-m-v} slices optimizer states on all ranks to decrease memory usage and removes corresponding \allgather.
Finally, line~\ref{line:adam-schedule:fuse-allreduce} fuses all operations in a single kernel.

\begin{figure}[t]
  \begin{subfigure}{\columnwidth}
  \centering
  \includegraphics[width=.85\linewidth]{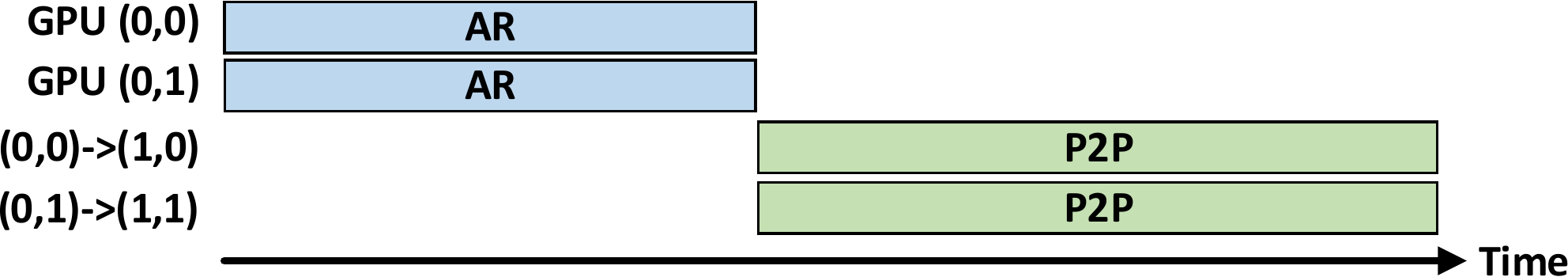}  
  \caption{In Megatron-LM each GPU sends redundant data. \label{fig:p2p-fusion-1}}
\end{subfigure}
\par\bigskip 
\begin{subfigure}{\columnwidth}
  \centering
  \includegraphics[width=.85\linewidth]{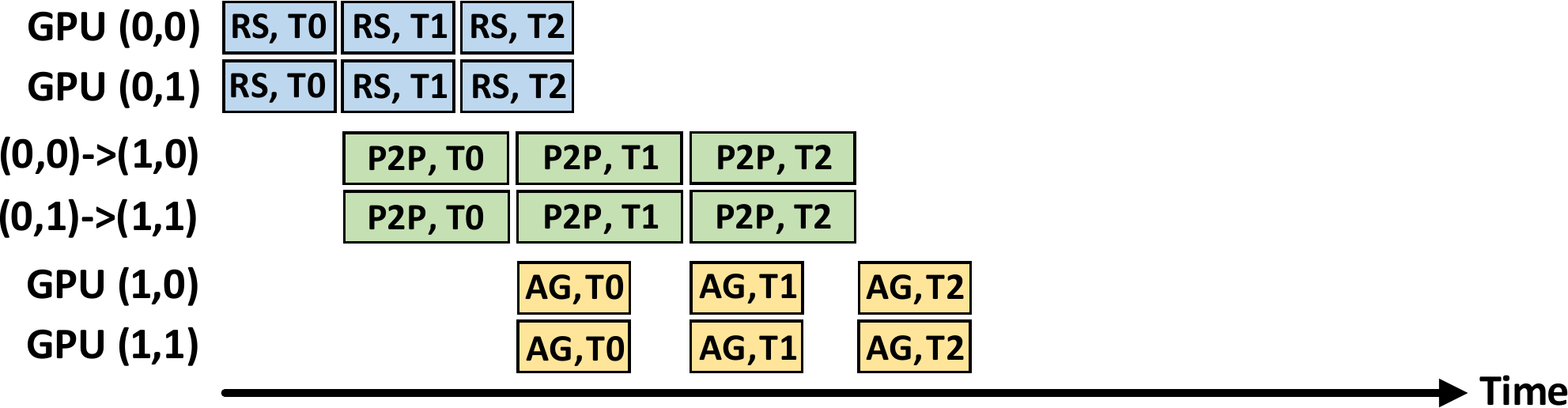} 
  \caption{Communication operations can be overlapped at the granularity of each 
  \emph{communication buffer tile} of data in single kernel 
  call.\label{fig:p2p-fusion-3}}
\end{subfigure}
  \caption{Two different schedules of pipeline parallelism.
\label{fig:P2Ptimeline}}
\end{figure}

\spara{Point-to-Point Communication in Pipeline Parallelism}: 
Figure~\ref{fig:p2p-fusion-1} shows a scenario of pipeline parallelism in Megatron-LM
with two transformer layers assigned to two groups each with two ranks. 
Rank $i$ in group $j$ is shown by $(j,i)$.
Each group uses model parallelism within its transformer layer.
Pipeline parallelism in Megatron-LM works as follows.
First, all ranks in the first group reduce their input using 
\allreduce to get replicated output.
Then each rank performs pointwise computations over the replicated output.
Finally, the first group sends the result of computations to the corresponding rank in the
second group using point-to-point (P2P) sends. (Line~\ref{line:p2p:comp2} in Figure~\ref{fig:traditional-p2p} shows these computations but are omitted in Figure~\ref{fig:P2Ptimeline} for simplicity). Since the output of \allreduce in 
Figure~\ref{fig:p2p-fusion-1} is replicated, redundant data is sent using P2P.
We can avoid this redundant communication by splitting
the \allreduce to \reducescatter and \allgather and reordering the P2Ps with the
\allgather. 
Hence, the inter-group communication is reduced by 
the group size.
We can further optimize by overlapping all communication operations. 
Figure~\ref{fig:p2p-fusion-3} shows that if the 
buffers are split into multiple tiles (\texttt{T0}--\texttt{T2} in the figure), 
intra-group and inter-group communications can be overlapped.

Figure~\ref{fig:traditional-p2p} is the original program, 
while Figure~\ref{fig:p2p-schedule} optimizes it by applying transformations.
Line~\ref{line:p2p:fuse-send} fuses the P2P 
send with computations.
Line~\ref{line:p2p:split} splits the \allreduce and reorders the returned \allgather with the fused P2P send at Line~\ref{line:p2p:reorder}.
Hence, P2P send and computations are performed 
on only a slice of data on the next group where the \allgather is also performed.
Finally, all three new operations get overlapped in Line~\ref{line:p2p:fuseAR}.

\begin{figure}[!t]
	\small
	\begin{subfigure}{\columnwidth}
		\begin{lstlisting}[language=DSL, numbers=left]
Var sum = AllReduce("+", in);
Var send = Dropout(recv+b,0.1) + r;|\label{line:p2p:comp1}||\label{line:p2p:comp2}|
Var output = Send(send, 
                  GroupRank(GROUP+1, RANK));

Execute transformer({in}, {output});
		\end{lstlisting}
		\caption{Traditional implementation. Each rank of a group sends same data to next group.}
		\label{fig:traditional-p2p}
	\end{subfigure}
	\par\bigskip 
	\begin{subfigure}{\columnwidth}
		\begin{lstlisting}[language=DSL, numbers=left]
fuseSend = fuse(send, output, SendFuse);|\label{line:p2p:fuse-send}|
(rsSum, agSum) = split(sum, ARSplitRSAG); |\label{line:p2p:split}|
(scSend, agOut) = reorder(fuseSend, agSum, 
                          AGReorder); |\label{line:p2p:reorder}|
overlapOut = overlap(rsSum, scSend, agOut); |\label{line:p2p:fuseAR}|
		\end{lstlisting}
		\caption{An Optimized Schedule. Each rank sends only a slice of data to ranks in next group and all operations are overlapped.}
		\label{fig:p2p-schedule}
	\end{subfigure}
	\caption{Optimizing pipeline parallelism of Megatron-LM. Input tensors: layer output \texttt{in}, bias \texttt{b}, and residual \texttt{r}. \label{fig:optimizing-p2p}}
\end{figure}

\section{The \tool Code Generator}
\label{sec:runtime}
\tool generates CUDA kernels for computation and communication operations for running on a distributed system with NVIDIA GPUs.
For each operation, \tool either generates (i) a call to a collective communication operation, 
(ii) a CUDA kernel for fused computations,
(iii) a CUDA kernel for fused-collective communications (Section~\ref{sec:code-gen-fused}), or
(iv) CUDA kernels for overlapping of communication and computation operations (Section~\ref{sec:overlap-impl}).
Moreover, \tool generates code for performing operations on multiple non-contiguous 
tensors (Section~\ref{sec:scattered-tensors}).
After generating CUDA kernels, \tool traverses the program's DFG to generate kernel calls.
\tool wraps generated programs as custom operators and integrates them into PyTorch, so that, applications like Megatron-LM can invoke them directly (Section~\ref{sec:pytorch-integration}).
We now discuss how \tool adapts NVIDIA Collective Communication Library (NCCL), a widely-used
hand-optimized high performance communication library, into a runtime
to execute above CUDA kernels. 

\subsection{NCCL Architecture}
\label{sec:nccl-arch}
NCCL communicates data stored in the global memory of one GPU to a memory location on another GPU using CUDA kernels.
NCCL's CUDA kernels perform communication by directly copying data from memory of one GPU to another GPU using GPUDirect Remote Data Memory Access~\cite{gpudirect}.
NCCL's architecture defines four key properties: (i) topology, (ii)
protocols, (iii) channels, and (iv) threads in a thread block of the
CUDA kernel. NCCL automatically sets key configuration values for these properties
based on the size of the input buffer, network architecture, and the size of
\WORLD. To ensure good performance, \tool's code generation must carefully reconfigure these properties
when extending NCCL to custom communication and computation.
We now provide a high level overview of these properties.

\spara{Topology} NCCL creates logical topologies, such as ring and tree, over the underlying interconnect network. 

\spara{Channels}
NCCL maps copies of a logical topology on the underlying interconnect network.
Each copy is called a channel and is assigned to one CUDA thread block.

\spara{Protocols}
NCCL sends data using one of the three protocols: \texttt{LL}, \texttt{LL128}, and \texttt{Simple}.
These protocols make different tradeoffs between latency and bandwidth based on the type of inter-node synchronization used: 
\texttt{LL} has the lowest latency and \texttt{Simple} provides the highest bandwidth.


\spara{Number of Threads}
NCCL sets a fixed number of threads for each channel (and thread block).
NCCL's kernels have high register usage, which limits the number of thread blocks per SM to one.

\spara{NCCL Workflow} 
After determining the topology, protocol, number of channels, and 
number of threads, NCCL calls its CUDA kernel for communication.
Each collective communication has three levels of tiling to fully utilize the massive parallelism of GPUs.
Data is first divided into \emph{buffer tiles} equal to the size of the communication buffer.
Each buffer tile is further divided among all ranks and channels to obtain \emph{chunks}.
Each channel communicates a chunk of data at a time.
The \emph{threads} in channels copy elements in and out of the buffers and
apply reduction operations (\texttt{sum}, \texttt{min}, \texttt{max}) if needed.
We now present details about \tool's code generation.

\subsection{Fused Collective Communications}
\label{sec:code-gen-fused}
Fused Collective Communication extends NCCL's existing kernels to enable arbitrary pointwise computations and reductions (i.e., beyond \texttt{min}, \texttt{max}, and \texttt{sum}).
We inspected more than 10K lines of code in NCCL to identify where computations can be added to pass intermediate values from communication to fused computations directly through registers.
\tool supports fusion of both pointwise operations and reductions into NCCL collectives.

Each NCCL protocol utilizes a different mechanism for communication and
\tool generates code for all of them. The important features of a protocol are
the pack type (64-bit for \texttt{LL}, 128-bit for \texttt{LL128} and \texttt{Simple}) and 
the load/store access pattern (shared memory for LL128, global memory for \texttt{LL} and \texttt{Simple}).
\tool generates template code for all element types in NCCL, and dispatches accordingly at runtime.
There are some subtleties in the code generation worth discussing:

\spara{Mixed Precision} 
When the element types of computations and the input tensors are different, \tool finds the largest element type and based on 
the pack type of the protocol calculates how many elements can be loaded at once.
All code will then be generated to operate on these many elements.

\spara{Sliced Tensor}
When a sliced tensor is used by a fused collective communication,
all memory accesses performed need to be mapped to elements of the sliced tensor.
\tool generates code that produces this mapping. 
To perform an \allgather on sliced tensors, the inverse of this mapping is
produced.

\spara{Tensor Reduction}
To reduce a \sliced tensor, each rank reduces locally and do an \allreduce.
This \allreduce reuses already established connections among ranks in the surrounding communication kernel to avoid extra startup latency.

\begin{figure*}[t]
	\centering
    \begin{subfigure}{0.48\linewidth}
    	\centering
    \includegraphics[width=\linewidth]{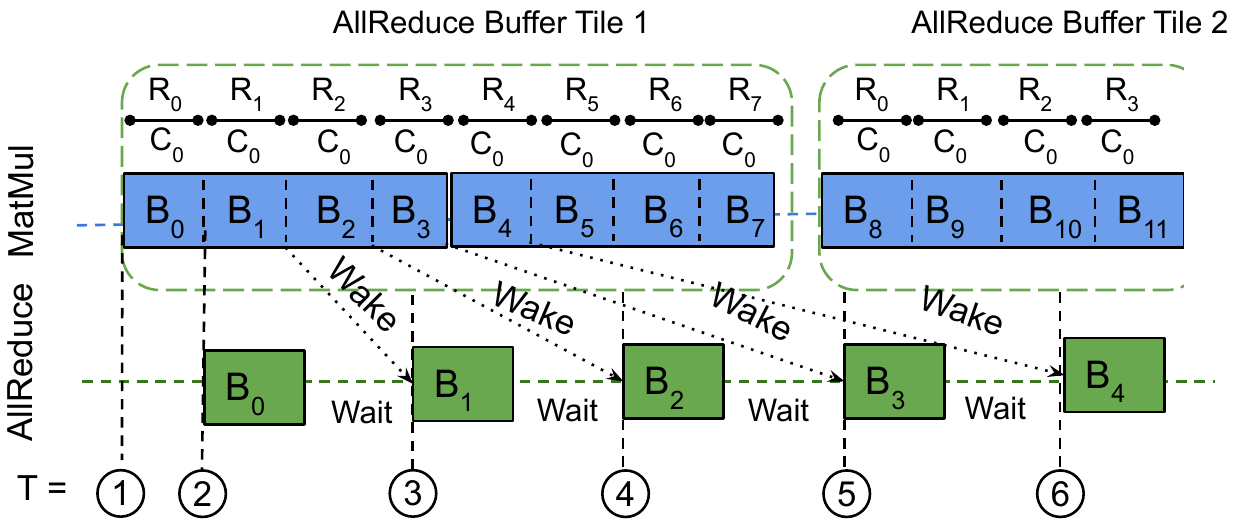}
    \caption{Workflow of overlap on \textbf{rank 0}. Rank 0 starts with chunk 0. }
	\end{subfigure}
    \hfill 
    \begin{subfigure}{0.48\linewidth}
    	\centering
        \includegraphics[width=\linewidth]{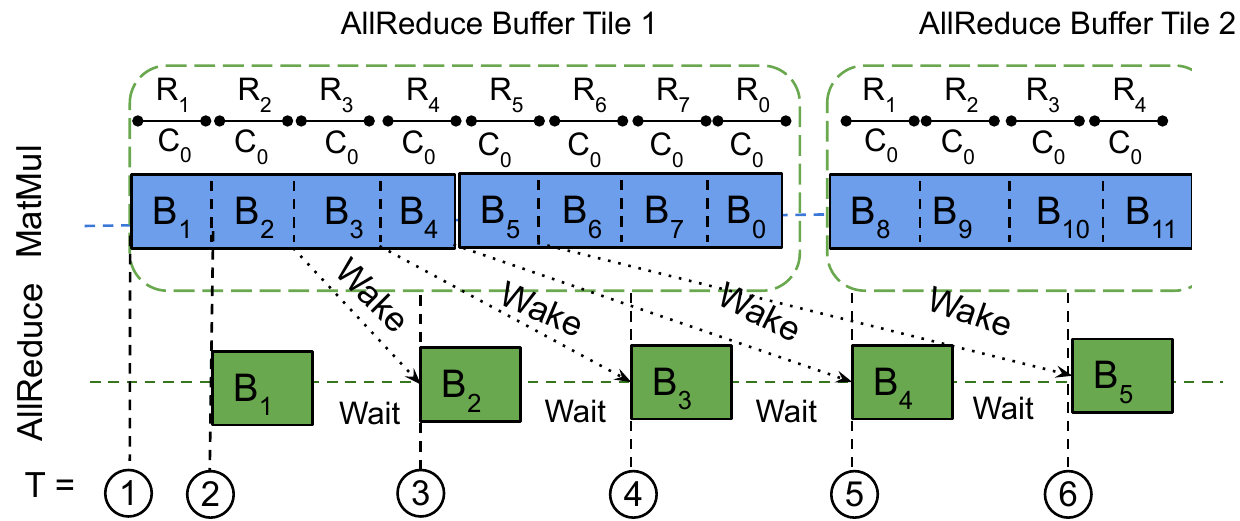}
        \caption{Workflow of overlap on \textbf{rank 1}. Rank 1 starts with chunk 1. }
    \end{subfigure}

	\caption{Workflow of \tool's overlapping of MatMul with \allreduce for a Float 16 matrix [8192, 3072] on 8 ranks (R$_0$ to R$_7$) with 1 channel (C$_0$) and 16 MB buffer size. Size of each 2-D chunk (B$_0$ to B$_{15}$) is [1024, 1024]. \tool's \allreduce and MatMul enables overlapping without decreasing the communication bandwidth and the efficiency of computation kernels.\label{fig:workflow-overlap}}
\end{figure*}

\subsection{Overlapping of Communication and Computation}
\label{sec:overlap-impl}
Overlapping of computation and communication has been studied in the context of executing stencil computations in a distributed system~\cite{Barigou2017,6799131,10.1145/2503210.2503289, 7573826,distributed-halide,KOZIRIS20031138,7336201,10.1145/1810085.1810091,8121995,10.1007/978-3-319-58667-0_18,sc20:pencil}.
These works use non-blocking MPI operations to communicate data and simultaneously perform computations on CPUs.
A similar approach for overlapping of computation and communication operations for a GPU workload would involve dividing all operations into sub-operations and ensuring dependency between sub-operations using CUDA streams.
However, this approach would provide sub-optimal performance because each sub-operation is performed on only a part of data, which leads to in-efficient computation and under-utilization of communication bandwidth.

Figure~\ref{fig:workflow-overlap} shows how the fine-grained overlapping of \tool addresses this issue using the example of a MatMul followed by a ring \allreduce.
First, it schedules the MatMul kernel (based on CUTLASS~\cite{cutlass})
to produce chunks in the same order as the \allreduce consumes them.
Here, the $n^{\text{th}}$ rank sends chunks in the order starting from the $n^{\text{th}}$ chunk. 
Hence, the MatMul kernel on $n^{\text{th}}$ rank produces chunks in the same order.
Second, \tool invokes both kernels only once on different streams and synchronizes the \allreduce with the MatMul using an efficient fine-grained spin-lock on a memory buffer to ensure that the \allreduce wakes up as soon as the MatMul produces a chunk.
Third, to provide opportunities to tune the 2-D tile sizes of the MatMul kernel, \tool generates a 2-D \allreduce kernel that communicates 2-D chunks, while NCCL \allreduce only supports 1-D continuous chunk.


	

The example in Figure~\ref{fig:workflow-overlap} works as follows.
At T = \circled{1}, all ranks invoke MatMul and \allreduce kernels.
On rank 0, after computing chunk 0, the MatMul kernel wakes the \allreduce kernel at T = \circled{2}, which starts communicating chunk 0.
While on rank 1, at T = \circled{2} the MatMul kernel wakes the \allreduce kernel to communicate chunk 1.
Concurrently, both MatMul kernels compute their corresponding next chunk.
At T = \circled{3}, MatMul kernels finished computing chunk 1 on rank 0 and chunk 2 on rank 1 and wakes up corresponding \allreduce kernels to communicate these chunks.
This process continues until all chunks are processed.

This process allows the MatMul kernel and \allreduce to be overlapped in a fine-grained manner,
which reduces the startup latency of \allreduce.
Since \allreduce communicates on the same chunk sizes, it achieves maximum communication bandwidth.
Furthermore, the MatMul kernel achieves maximum efficiency because the kernel is invoked on the full matrix size.
Figure~\ref{fig:matmul-overlap-intro} shows that this overlapping provides up to 1.36$\times$ better performance and hides more than 80\% of the MatMul time.

\subsection{Operations on Scattered Tensors}
\label{sec:scattered-tensors}
In data parallelism, communication and computation occur on different layers of widely different sizes.
Since machine learning frameworks allocate parameters and gradients of layers in non-contiguous buffers,
gradients are copied to a large buffer to avoid launching multiple \allreduce operations.

\tool supports generating a single kernel for both computation and communication operations acting on non-contiguous tensors.
In this section, we show how \tool modifies NCCL to generate a single communication kernel for scattered tensors.
This code generation is non-trivial because NCCL has several design decisions based on the assumption that it is communicating a single contiguous buffer.
For example, each thread of a NCCL channel copies only a few elements in each iteration, and hence indexing the correct tensor at a particular offset requires a linear search through all non-contiguous tensors, which can lead to significant overhead.
\tool solves this problem by first dividing each tensor into buckets of size at most 2$^{10}$ elements and then assigning buckets to warps in a round-robin manner.
This mechanism allows each thread to quickly find the offset in a tensor, since a warp can directly index in its assigned bucket.
\tool pre-calculates the number of buckets that belong to the same contiguous buffer and calculates the offset for all of them once.

The process of breaking each tensor to buckets has computation overhead and extra memory requirements.
Since this bucketing is done only once on the CPU and training tasks run for thousands of iterations on the same tensors, the computation overhead is negligible.
Each bucket is represented by a pair of 64-bit tensor address and a 32-bit offset into the associated tensor, leading to $12 \times \left \lceil \frac{N}{2^{10}}\right \rceil$ bytes of extra memory for a tensor with $N$ elements.
However, this memory overhead is negligible for large models. For example, for BERT model with 334M elements, the memory requirement is 0.6\%.
Table~\ref{tab:scattered-tensors} shows that the overhead of scattered tensors is insignificant over contiguous tensors.

\begin{table}[t]
    \center
  \small
  \caption{Time to perform parameter update of all 360 tensors of BERT using Adam/LAMB on 256 Tesla V100 GPUs with scattered tensors implementation and a single contiguous tensor of size equal to the sum of size of all tensors.\label{tab:scattered-tensors}}
  \begin{tabular}{c|c|l}
  \textbf{Optimizer}   & \textbf{Scattered Tensor} & \textbf{Single Tensor}\\ \hline
  Adam & 33.89 ms & 33.21 ms  \\ 
  LAMB & 37.04 ms & 36.71 ms
  \end{tabular}
  \par \bigskip
\end{table}

\subsection{PyTorch Integration}
\label{sec:pytorch-integration}
We integrated \tool generated code as a function to PyTorch's \texttt{torch.distributed} module.
This design allows us to re-use the logic for initializing NCCL and provide compatibility with models already using \texttt{torch.distributed}.
We added wrapper functions for calling \tool generated operations.
These wrapper functions prepare the arguments for calling \tool's operations, which includes pre-calculating pointers to the buckets for scattered tensors and clearing the spin-lock buffers for overlapping.
Machine learning models can invoke \tool functions using PyTorch.
\section{Evaluation}
\label{sec:experiments}

This section evaluates the effectiveness of \tool through standalone experiments and end-to-end distributed machine learning scenarios of data, model, and pipeline parallelism.

Our experiments are performed on a cluster of 16 NVIDIA DGX-2 nodes where each node contains dual 24-core Intel Xeon CPUs and 16 NVIDIA Tesla V100 (32GB) GPUs.
Each GPU within a node is connected to six NVSwitches with six NVLinks (25~GBps per NVLink).
Nodes are connected with 8 non-blocking EDR InfiniBand (100~Gbps) network.
All nodes run Ubuntu 20.04, CUDA~11.3, cuDNN 8.2 and PyTorch 1.10.

\subsection{Data Parallel Training}
In data parallelism, communication involves an \allreduce of gradients among all ranks. The output is used by the optimizer to update the model parameters.  We evaluate \tool generated code for two widely-used optimizers, Adam and LAMB.
All our experiments in this section were performed on all 16 DGX-2 nodes in our cluster.

\subsubsection{Standalone Experiments}
We first perform standalone experiments to explore different \tool schedules over a range of input tensors from $2^{10}$ to $2^{30}$ elements.
The autotuner generates and executes implementations with different configurations, including all NCCL protocols and all channels from 2 to 64.
For each tensor, the autotuner reports the best average result of 1000 iterations.

\spara{Baselines} The baselines perform parameter update by first doing \allreduce over gradients and then call FusedAdam or FusedLAMB from NVIDIA Apex~\cite{apex}. 
Both FusedAdam and FusedLAMB fuses all the parameter update computations.

\spara{\tool Schedules}
The autotuner generates following three schedules of Adam and LAMB by applying different \tool transformations for each input size and reports the best schedule to the user for each input size:
\begin{enumerate}[leftmargin=*,topsep=2pt]
  \item \textbf{AR-Opt} (Opt = Adam/LAMB) refer to the traditional parameter update technique, i.e., an \allreduce over gradients  and then each GPU individually performs the optimizer computation. These schedules fuse all computations into a single kernel, thereby simulating the baseline implementations of FusedAdam and FusedLAMB.
  \item \textbf{GShard-Eq} or \textbf{RS-Opt-AG} (Opt = Adam/LAMB) are generated from \textit{AR-Opt} by first splitting the \allreduce into \reducescatter and \allgather, and then reordering \allgather with the fused optimizer computations.
  Hence, these schedules distribute parameter update across all ranks, similar to GShard~\cite{gshard} and ZeRO~\cite{zero}. Since GShard does not support execution on GPUs, we refer to this schedule as GShard-Eq in our results. 
  \item \textbf{fuse(RS-Opt-AG)} (Opt = Adam/LAMB) are generated by fusing all operations of \textit{RS-Opt-AG} into FusedAllReduce.
\end{enumerate}

\paragraph{Results}
\begin{figure}[t]
	\centering
    \begin{subfigure}{0.93\linewidth}
    \centering
    \includegraphics[width=\linewidth]{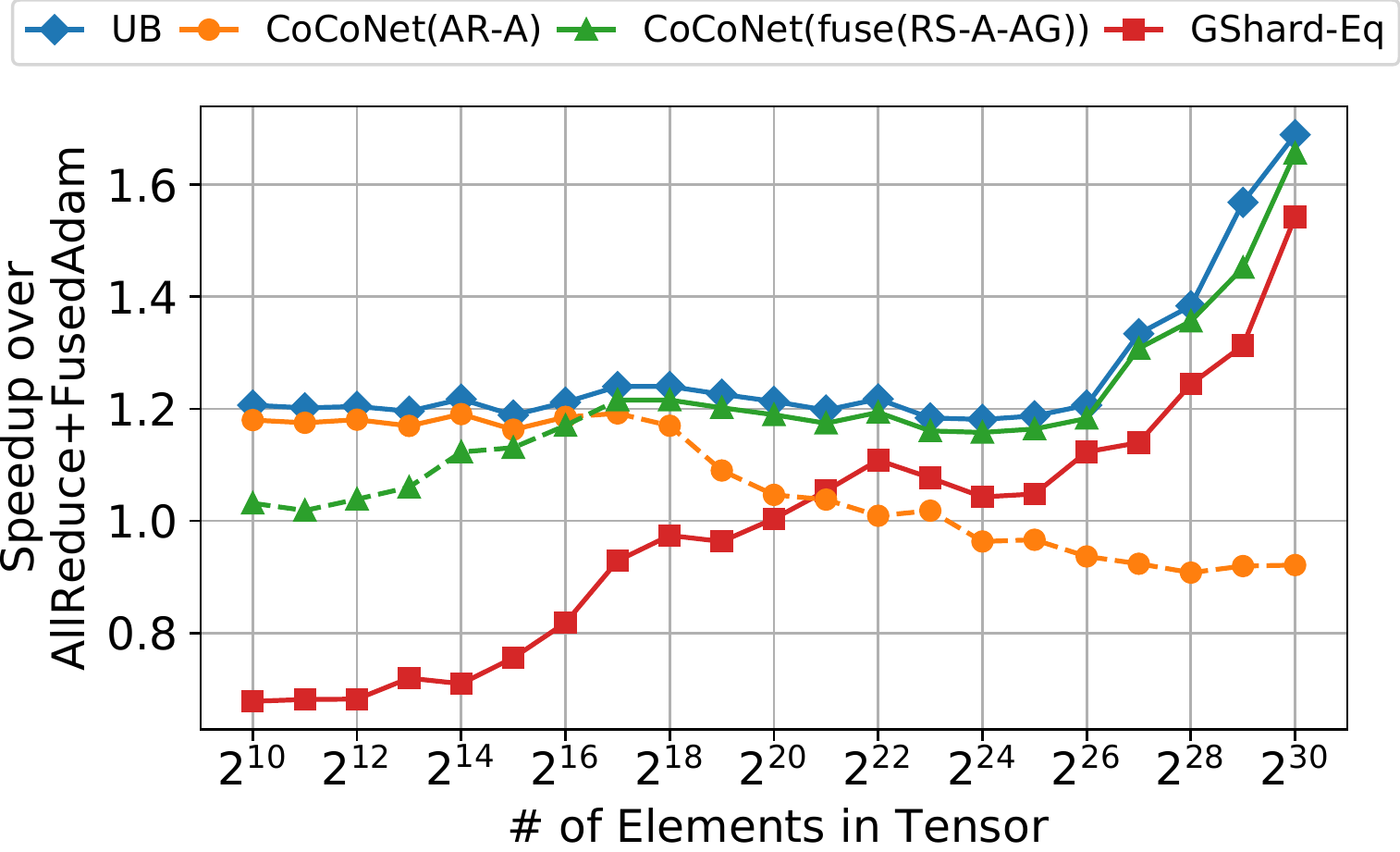}  
    \caption{Mixed-precision Adam. AR-Adam(AR-A) runs best till 2$^{16}$. fuse(RS-A-AG) represents fuse(RS-Adam-AG) and runs best after 2$^{17}$.}
    \label{fig:bandwidth64GPUs:adam}
  \end{subfigure}
  \par \bigskip
  \begin{subfigure}{0.93\linewidth}
    \centering
    \includegraphics[width=\linewidth]{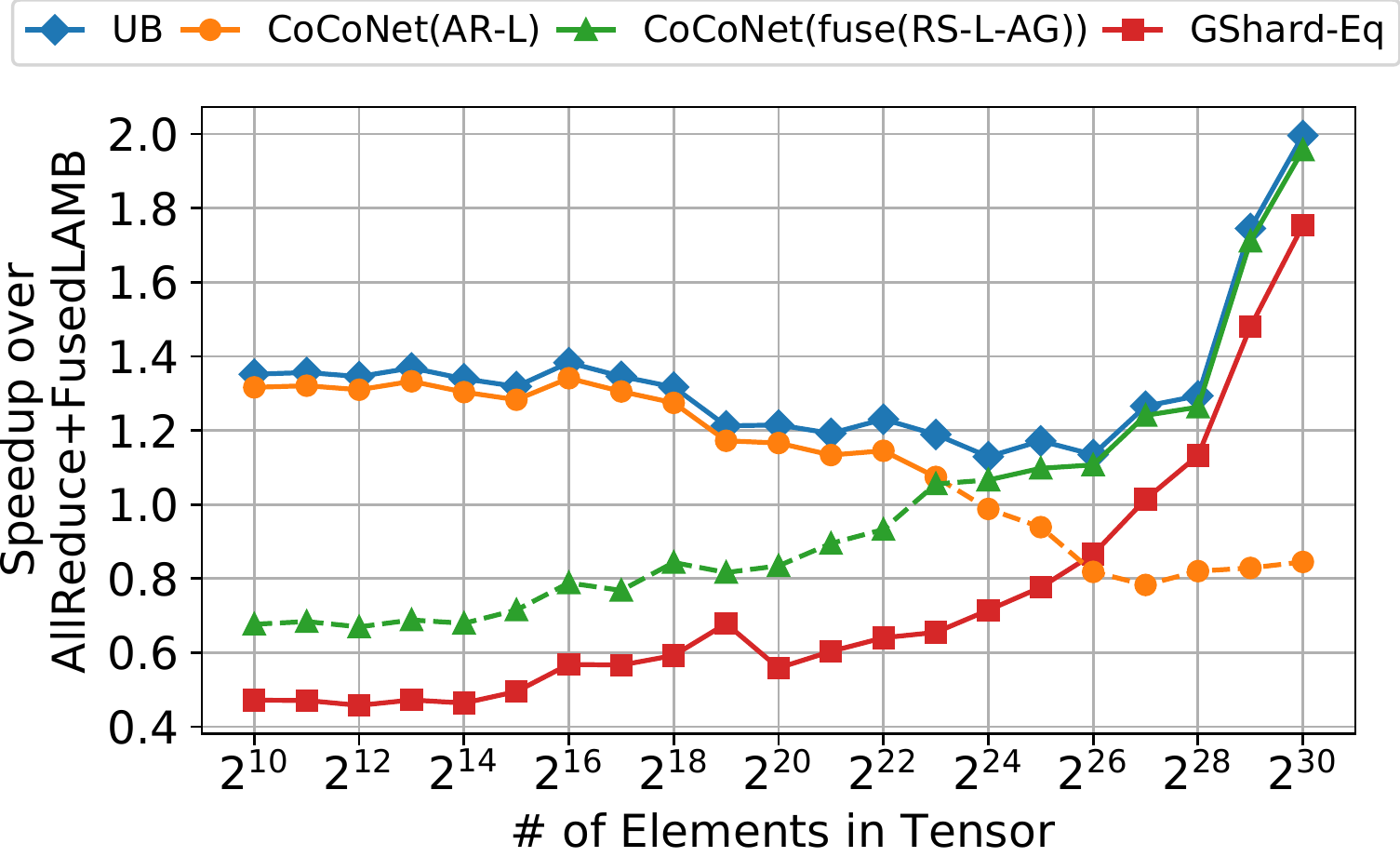}  
    \caption{Mixed-precision LAMB. AR-LAMB(AR-L) runs best till 2$^{16}$. fuse(RS-L-AG) represents fuse(RS-LAMB-AG) and runs best after 2$^{17}$.} 
    \label{fig:bandwidth64GPUs:lamb}
  \end{subfigure}
  \caption{\tool speedup on 256 GPUs. For each size, \tool chooses the best schedules.
   UB (upper bound) takes AllReduce-only as max achievable speedup.
   }
  \label{fig:bandwidth64GPUs}
\end{figure}

Figure~\ref{fig:bandwidth64GPUs} shows the speedup of \tool schedules over the baseline
for several tensor sizes. 
The results are shown for mixed-precision~\cite{mixed-precision-training} using Float 16, and the results for Float 32 are qualitatively similar. 
In these figures, UB represents the cost of \allreduce alone without doing any computation, and thus is the upper bound of possible speedups. 

Even though the \textit{AR-Opt} schedules emulate the baseline implementations, they are faster on smaller tensors.
This is because the baseline implementations perform additional preprocessing to optimize the amount of thread-parallelism and instruction-level parallelism per invocation. While this preprocessing cost hurts smaller tensors, its benefit shows up for larger tensors where \textit{AR-Opt} performs worse. 

Since \textit{GShard-Eq} and \textit{fuse(RS-Opt-AG)} schedules distribute the optimizer computation,  they perform better than the baseline for large tensors.
The performance of \textit{fuse(RS-Opt-AG)} shows the advantage of fusing computation and communication kernels as these schedules achieve near optimal speedups for large tensors. 
These schedules are respectively 13\% and 14\% faster than GShard-Eq for Adam and LAMB.

For smaller tensor sizes, multiple kernel calls are required for GShard-Eq schedules significantly hurt performance. Interestingly, \textit{fuse(RS-Opt-AG)} schedules are slower than \textit{AR-Opt} schedules for smaller tensor sizes though they require one less kernel call because the fused kernels have a higher register usage, thereby restricting the thread-level parallelism. This demonstrates that the fusion of communication and computation is not always a good idea.

Table~\ref{tab:loc:data-parallel} shows that the lines of generated code for each schedule are significantly more than the implementation in \tool and the autotuner explored all schedules in 10 seconds.
In summary, \tool provides performance improvements over baselines with fewer lines of code.
The \textit{AR-Opt} and the \textit{fuse(RS-Opt-AG)} reach close to optimal performance for smaller and larger tensors respectively. This amounts to a speedup of 1.2$\times$ to 1.7$\times$ for Adam and 1.35$\times$ to 2.0$\times$ for LAMB.  
There is no schedule that performs best for all sizes, which demonstrates the need for the autotuner.

\begin{table}
  \centering
	\small
  \caption{Lines of code of implementation of distributed machine learning workloads in CUDA and \tool{}, and time taken by the autotuner to find the best schedule.}
  \label{tab:loc-autotuner-time}

  \begin{subtable}{\linewidth}
    \centering
    \caption{Data Parallel optimizer update using Adam and LAMB\label{tab:loc:data-parallel}}
  \begin{tabular}{lrrr}
    \textbf{Schedule}        & \textbf{\thead{Generated\\CUDA}} & \textbf{\thead{Program in\\\tool{}}} & \textbf{\thead{Autotuner\\Time}}\\ \hline
    \textit{AR-Adam}           & 16   &  12  & \multirow{3}{*}{9 secs}    \\
    \textit{RS-Adam-AG}        & 24   &  16  &  \\
  \textit{fuse(RS-Adam-AG)}  & 150  &  17  & \\\hline
  \textit{AR-LAMB}           & 80   &  15  & \multirow{3}{*}{10 secs}   \\
  \textit{RS-LAMB-AG}        & 140  &  17  &   \\
  \textit{fuse(RS-LAMB-AG)}  & 220  &  18  & \\
  \hline
  \end{tabular}
  \par \bigskip
  \end{subtable}

  \begin{subtable}{\linewidth}
    \centering
    \caption{Model Parallel Self Attention and Multi Layer Perceptron\label{tab:loc:model-parallel}}
  \begin{tabular}{lrrr}
    \textbf{Schedule}       & \textbf{\thead{Generated\\CUDA}} & \textbf{\thead{Program in\\\tool{}}} & \textbf{\thead{Autotuner\\Time}}\\ \hline
    \textit{MM-AR-C}               &  20  &  10   & \multirow{3}{*}{12 secs} \\
  \textit{MM-RS-C-AG}            &  140 &  13    & \\
  \textit{ol(MM,fuse(RS-C-AG))}  &$\approx$ 2k    &   14  &   \\\hline
  \end{tabular}
  \par \bigskip
  \end{subtable}

  \begin{subtable}{\linewidth}
    \centering
    \caption{Pipeline Parallel Transformer Layer\label{tab:loc:pipeline-parallel}}
  \begin{tabular}{lrrr}
    \textbf{Schedule} & \textbf{\thead{Generated\\CUDA}} & \textbf{\thead{Program in\\\tool{}}} & \textbf{\thead{Autotuner\\Time}}\\ \hline
    \textit{AR-P2P-C-AG}               &  20  &  10  & \multirow{3}{*}{11 secs}  \\
    \textit{RS-P2P-C-AG}            &  140 &  13    & \\
    \textit{ol(RS,fuse(P2P-C),AG))}  &$\approx$ 2k    &   14  &   \\\hline
  \end{tabular}
  \par \bigskip
  \end{subtable}
\end{table}

\subsubsection{Integeration with BERT}
\label{sec:experiments:bert}
We use \tool generated optimizers to train three large BERT models from NVIDIA~\cite{nvbert}.
We use mixed precision training with both Adam with 8192 global batch size and LAMB with 65536 global batch size.

\spara{Baselines} We consider three baselines for this experiment:
\begin{itemize}
  \item \textbf{NV BERT}~\cite{nvbert} is the NVIDIA BERT Script. 
  It copies gradients of each layer into a single buffer, calls \allreduce on the buffer, and
  copy back the results into original gradients. Finally, it calls either FusedAdam or FusedLAMB.
  \item \textbf{PyTorch DDP}~\cite{pytorch-ddp} stores all gradients in buckets of 25MB and overlaps the \allreduce on each gradient bucket with computations during training. After reducing all gradients it calls FusedAdam or FusedLAMB.
  \item \textbf{ZeRO}~\cite{zero}
  copies gradients into a contiguous buffer and then distributes Adam's computation similar to \textit{RS-Opt-AG} schedules above.
  The ZeRO implementation of LAMB does not support distributing optimizer state among GPUs because significant engineering efforts are required to implement reduction over distributed gradients and weights in a distributed LAMB implementation~\cite{deepspeed490}.
\end{itemize}

\spara{\tool Integeration} We integrated the scattered tensors implementation of \textit{fuse(RS-Opt-AG)} schedule for both Adam and LAMB in PyTorch. 
These implementations provide three benefits over the baselines: (i) the scattered tensor implementation avoids copying all gradients to a single buffer and allocating this buffer, (ii) the fused schedule performs best for the tensor sizes used in BERT, and (iii) the fused schedule distributes memory of optimizer state among all GPUs.

\begin{table*}
	\small
  \centering
  \caption{Maximum Micro Batch Size supported by all implementations and speedup of \tool over the baselines when training BERT with three parameter configurations using Adam and LAMB optimizer. OOM represents Out of Memory.\label{tab:bert-results}}
  \begin{tabular}{cccccccccccc}
  \textbf{Optimizer} & \textbf{\# of Parameters} & \multicolumn{4}{c}{\textbf{Maximum Micro Batch Size}} & \multicolumn{3}{c}{\textbf{Speedup of \tool over}}\\
  \cmidrule(lr){3-6} \cmidrule(lr){7-9}
                          & & NV BERT & PyTorch DDP & ZeRO & \tool & NV BERT & PyTorch DDP & ZeRO  \\
  \hline
  \multirow{3}{*}{Adam} & 336 M & 32  &  32  & 32 & 32 & 1.18$\times$ & 1.22$\times$ & 1.10$\times$\\ 
  &1.2 B & 8   &  8   & 32 & 32 & 1.53$\times$ & 1.52$\times$ & 1.10$\times$\\ 
  &3.9 B & OOM & OOM & 8  & 8 & -- & -- & 1.22$\times$\\
  \hline
  \multirow{3}{*}{LAMB} & 336M & 64 & 64 & 64 &128 & 1.20$\times$ & 1.20$\times$ & 1.15$\times$\\
  & 1.2B & 8 & 8 & 8 & 64 &1.67$\times$ & 1.68$\times$ & 1.64$\times$\\ 
  & 3.9B & OOM & OOM & OOM & 8 & -- & -- & --\\ 
  \hline
  \end{tabular}
  \par \bigskip
\end{table*}

\spara{Results} Table~\ref{tab:bert-results} shows the speedup provided by \tool in training three BERT models over baselines.
For Adam optimizer, \tool provides speedup over all baselines in training BERT 336M because \tool's fused schedules perform better than other implementations.
\tool provides even higher speedup on larger BERT models because the fused schedules decrease memory usage by distributing Adam's state over all GPUs, which improves the efficiency of matrix multiplication GPU kernels by enabling higher batch size per iteration.
For example, for BERT 1.2B \tool provides 1.53$\times$ speedup over NV BERT and PyTorchDDP because of the optimized fused schedule and higher batch size enabled by \tool.
On 3.9B parameter model, NV BERT and PyTorch go Out of Memory.
ZeRO also supports higher batch size for BERT 1.2B and 3.9B but \tool still gives speedup because of the advantages of scattered tensor implementation of fused schedules.

Results for LAMB are similar. 
\tool provides up to 1.64$\times$ speedup over all baselines. 
For LAMB, the speedup over ZeRO is higher than Adam because ZeRO does not support distributing LAMB optimizer state, and hence, supports smaller batch sizes as compared to \tool.

In summary, \tool significantly improves data-parallel training time of BERT models. 
\tool's schedules can be automatically generated and \tool's scattered tensors implementation can support a  wide range of optimizers.
Not only does the fusion of computation and communication lead to performance improvement over the baselines of PyTorch DDP and ZeRO, it also decreases the memory usage, which helps in increasing the batch size to train models faster.

\subsection{Model Parallelism}
Megatron-LM~\cite{megatronlm} uses a model parallel approach for inference and training of transformer models, such as BERT~\cite{bert} and GPT-2~\cite{gpt-2}. 
A transformer layer contains a self-attention block and a 
multi-layer perceptron (MLP) block. Last few operations of a self-attention block are the same 
computations as shown in Figure~\ref{fig:traditional-mp}. An MLP block's last operations are similar to 
Figure~\ref{fig:traditional-mp} with the input tensor and weight sizes as $[B, S, 4 \times H]$ and $[4 \times H, H]$ ($B$, $S$, and $H$ are batch size, sequence length, and hidden size, respectively).
Since model parallelism is applied within one node, all experiments in this section are performed on a single NVIDIA DGX-2 node.

\subsubsection{Standalone Experiments}
We first perform standalone experiments to evaluate different schedules generated by the autotuner.
We compare following schedules for model parallel self-attention code of Figure~\ref{fig:traditional-mp} and similar operations of multi-layer perceptron:
\begin{enumerate}[leftmargin=*,topsep=2pt]
  \item \textbf{Megatron-LM} is the baseline implementation of Figure~\ref{fig:traditional-mp} in Megatron-LM.
  \item \textbf{MM-AR-C} improves the Megatron-LM implementation by fusing all pointwise computations into one kernel.
  \item \textbf{GShard-Eq} or \textbf{MM-RS-C-AG} uses the same techniques as GShard. It is generated from \textit{MM-AR-C} by splitting the \allreduce into a \reducescatter and an \allgather, and reorders \allgather with computations. 
  This schedule represents GShard because GShard is not available for GPUs.
  \item \textbf{ol(MM,fuse(RS-C-AG)} is generated from the previous schedule by fusing the \reducescatter, computation, and \allgather into a FusedAllReduce and then overlapping it with the MatMul. 
  The autotuner returned this as the best schedule and hence represents \tool in our results.
\end{enumerate}
\begin{figure}[t]
	\centering
  \includegraphics[width=\linewidth]{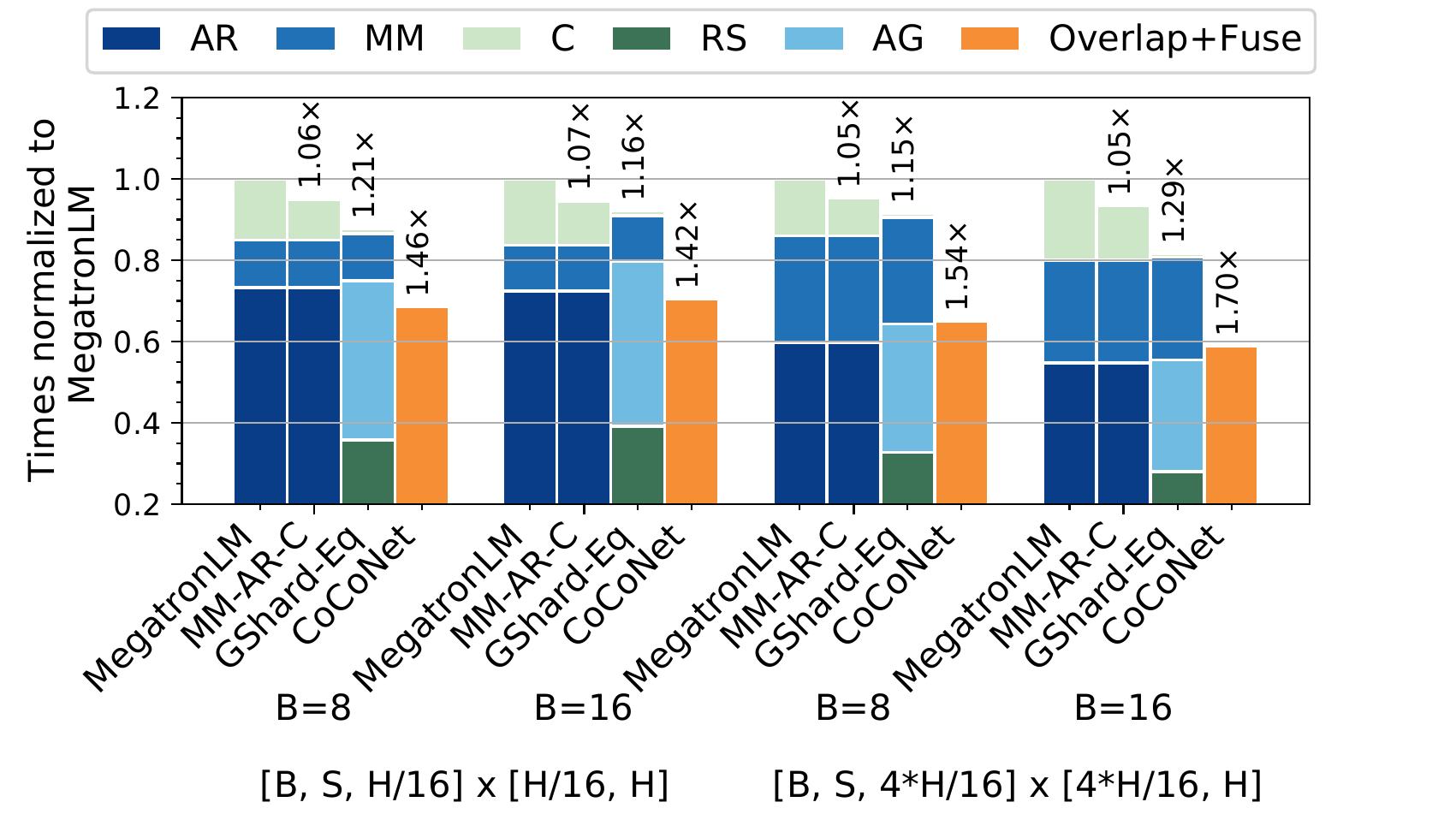}
  \caption{Times of \tool's schedules of model parallel self-attention and multi-layer perceptron of GPT-2 normalized to corresponding Megatron-LM's implementation.
  \label{fig:matmul-overlap}}
\end{figure}

\spara{Results} We evaluate these schedules with sizes of GPT-2 8.3 Billion parameter model (i.e., $S=1024$, $H=3072$) for 8 and 16 batch sizes.
Figure~\ref{fig:matmul-overlap} shows the times of all schedules normalized to the time of implementation in Megatron-LM.
\textit{MM-AR-C} schedule provides speedup over Megatron-LM's implementation because this schedule fuses all pointwise computations in a single GPU kernel.
GShard-Eq (\textit{MM-RS-C-AG}) provides 1.15$\times$ to 1.29$\times$ speedup over Megatron-LM by distributing computations on all ranks.
\tool's best schedule (\textit{ol(MM,fuse(RS-C-AG))}) provides 1.42$\times$ to 1.70$\times$ speedup over Megatron-LM and 1.21$\times$ to 1.34$\times$ over GShard-Eq because it overlaps Fused\allreduce with the matrix multiplication.
Table~\ref{tab:loc:model-parallel} shows that the lines of generated CUDA code for each schedule are significantly more than the implementation in \tool and the autotuner explored all schedules in 12 seconds.

\subsubsection{Integration with Megatron-LM}
\label{sec:experiments:model-parallel:integeration}
After integrating \tool's overlap schedule in Megatron-LM,
we found that \tool improved inference times of BERT 3.9B parameter model by 1.51$\times$ and GPT-2 8.3B parameter model by 1.48$\times$.
Hence, overlapping matrix multiplication with fused collective communication significantly improves inference times.

\subsection{Pipeline Parallelism}
\tool can decrease inference times in pipeline parallelism by fusing computation and communication and overlapping multiple communication operations.
We evaluate \tool on computations of model and pipeline parallelism in Megatron-LM for GPT-2 8.3B and GPT-3 175B parameter models.
A transformer layer contains several operations but the operations of interest for this experiment are presented in Figure~\ref{fig:traditional-p2p}.
All experiments in this section are performed on all 16 NVIDIA DGX-2 nodes.

\subsubsection{Standalone Experiments}
We first perform standalone experiments to evaluate different schedules generated by the autotuner.
We compare the following schedules for pipeline parallelism code of Figure~\ref{fig:traditional-p2p}:
\begin{enumerate}[leftmargin=*,topsep=2pt]
 \item \textbf{Megatron-LM} is the implementation of Figure~\ref{fig:traditional-p2p} in Megatron-LM and serves as a baseline for this experiment.
 \item \textbf{AR-C-P2P-AG} is generated by slicing the output of \allreduce to perform sliced P2P sends and computations, and finally an \allgather to collect the output of computations.
 This schedule improves over Megatron-LM by slicing the P2P sends and fusing all the computations.
 \item \textbf{GShard-Eq} or \textbf{RS-C-P2P-AG} is generated from the previous schedule by splitting the \allreduce into a \reducescatter and an \allgather, then reordering the \allgather with P2P send and computations.
 Since this schedule is similar to GShard, it represents GShard-Eq in our results.
  \item \textbf{ol(RS,fuse(C-P2P),AG)} is generated from previous schedule by fusing computations with P2P sends, and overlapping all three communication operations (Figure~\ref{fig:p2p-fusion-3}). This schedule is returned by the autotuner as the best schedule and hence, represents \tool in our results.
\end{enumerate}

\begin{figure}[t]
	\centering
  \includegraphics[width=1\linewidth]{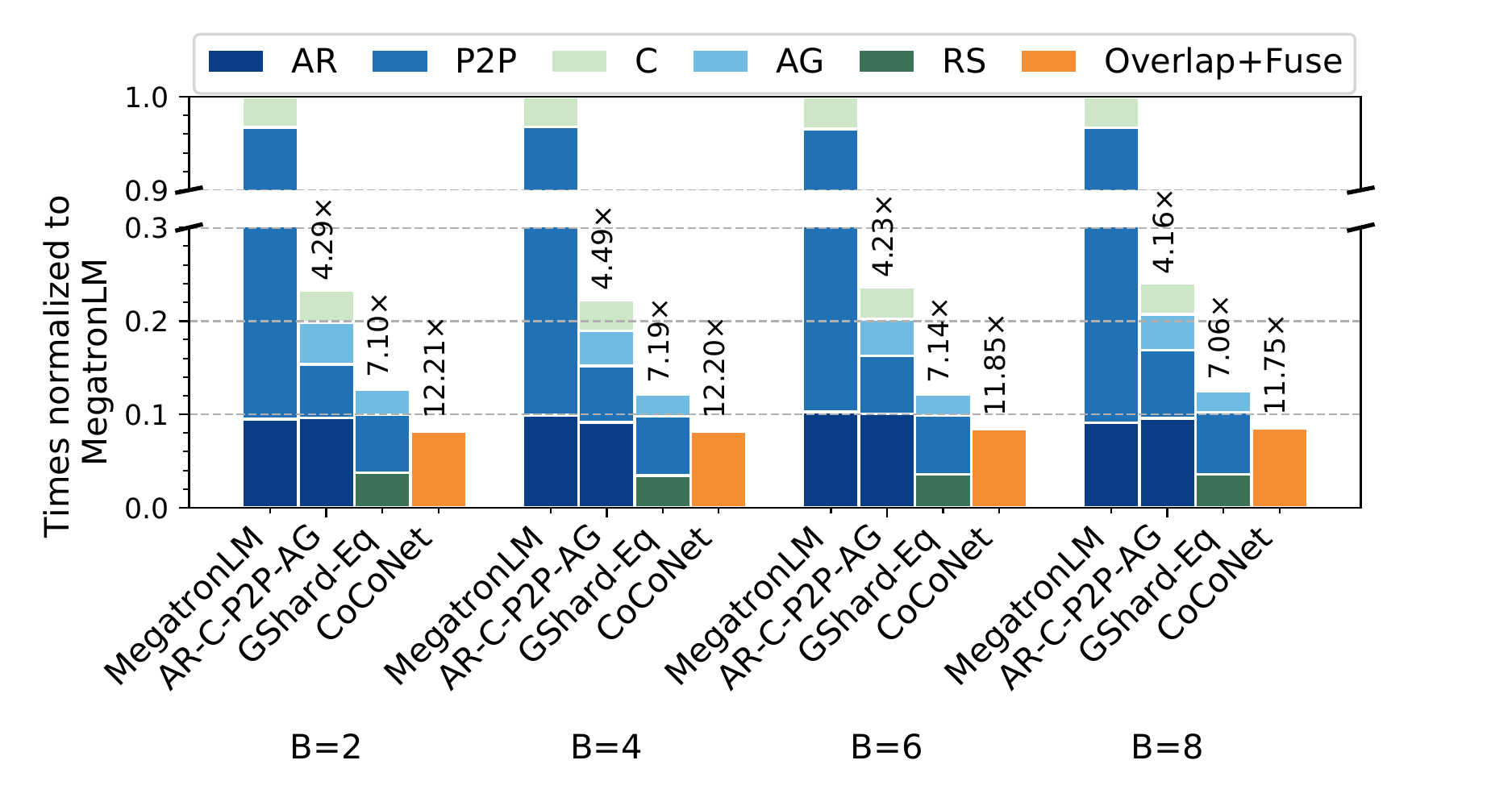}
  \caption{Times of three schedules for GPT-3 175B in \tool for pipeline and model parallelism normalized to Megatron-LM's corresponding implementation. \label{fig:pipeline-overlap}}
\end{figure}

\spara{Results} 
Figure~\ref{fig:pipeline-overlap} shows the breakdown of each operation
with one transformer layer assigned to each node. The sequence length ($S=2048$) and the hidden
size ($H=12288$) are of GPT-3 175B model.
\tool's best schedule \textit{ol(RS,fuse(C-P2P),AG)} is 11.75$\times$--12.21$\times$ faster than Megatron-LM's implementation, 2.84$\times$ faster than \textit{AR-C-P2P-AG}, and 1.66$\times$--1.72$\times$ faster than GShard (\textit{RS-C-P2P-AG}). 
The speedups are because: 
(i) sliced P2P reduces cross node communication volume, 
(ii) fusing communication and computation operations improves memory bandwidth utilization, and
(iii) overlapping communication using different connections 
(NVLink within node and InfiniBand across nodes) improves network bandwidth utilization, while other schedules utilize only one stack at a time.
Table~\ref{tab:loc:pipeline-parallel} shows that the lines of generated CUDA code for each schedule are significantly more than the implementation in \tool and the autotuner explored all schedules in 11 seconds.

\subsubsection{Integration with Megatron-LM}
\label{sec:experiments:pipeline-parallel:integeration}
We evaluated the inference throughput of GPT-2 8.3B and GPT-3 175B parameter models by integrating \tool's \textit{ol(RS,fuse(C-P2P),AG)} schedule in Megatron-LM.
Table~\ref{tab:pipeline-results} shows the speedups achieved by \tool.

\tool significantly improves inference throughput of GPT-3 and GPT-2 due to its fusion and fine-grained overlapping of multiple communication operations.
\begin{table}[t]
  \caption{Speedup in inference by \tool's implementation of pipeline parallelism for GPT-2 and GPT-3.
  Layers per node were obtained by equally distributing layers on all nodes.
  To evenly distribute layers of GPT-2, number of layers were increased to the nearest multiple of 16, i.e., 80.
  \label{tab:pipeline-results}}
  \begin{tabular}{lcccc}
    Model & \thead{Layers \\per node} & \thead{Maximum \\ Micro Batch Size} & Speedup \\
    \midrule
    GPT-2 8.3B & 5 & 16 & 1.77$\times$\\
    GPT-3 175B & 6 & 2 & 1.33$\times$\\
  \end{tabular} 
  \par \bigskip
  \end{table}
\section{Related Work}
\label{sec:related}

\spara{Distributed Machine Learning Abstractions}
Existing machine learning frameworks~\cite{mxnet,tensorflow,jia2014caffe,pytorch,sergeev2018horovod} and DSLs~\cite{tvm18,distributed-halide} provide abstractions for writing distributed machine learning workloads.
Similar to \tool, in these abstractions, a distributed machine learning program takes input tensors, performs operations on tensors, and returns tensors as the output.
However, unlike these abstractions, \tool preserves the layout information for each tensor.
The layout information enables \tool to perform static type checking of each operation, and automatically perform transformations on the program, which is not possible with existing abstractions.

\spara{Distributed Neural Network Training} 
Several works have improved data-, model-, and pipeline-parallel techniques for both training and inference.
Mesh-Tensorflow \cite{meshtensorflow} and GShard~\cite{gshard} create \emph{shards} of weights and model state that can be split among ranks.
Horovod~\cite{sergeev2018horovod} introduced the \emph{Tensor Fusion} optimization that copies all gradients to a single buffer of 64MB, calls \allreduce on the buffer, and then copies the updated value to original gradients.
ZeRO~\cite{zero} splits weights and model state among ranks and uses \reducescatter and \allgather to distribute computation.
FlexFlow~\cite{flexflow} performs operator splitting as a way to represent both data-parallelism and model-parallelism, but does not optimize computation with communication.
\tool provides several optimizations over these works that are possible only by breaking the abstraction: (i) scattered tensors that remove extra storage and memory copy operations, (ii) fusion communication collectives, and (iii) novel communication and computation overlapping techniques.
PyTorch's DDP~\cite{pytorch-ddp} overlaps \allreduce of gradients with the forward and backward pass.
However, unlike \tool, PyTorch's DDP requires extra memory for overlapping, which can increase training time for very large models~\cite{megatronlm-github} and do not support slicing of optimizer parameter update that significantly decrease memory usage.
GPipe~\cite{gpipe}, Pipedream~\cite{pipedream}, and Narayanan et al.~\cite{narayanan2021efficient} proposed pipeline training to improve model parallelism, by dividing the forward and backward pass into several mini-batches, which are then pipelined across devices.
vPipe~\cite{vpipe} improves these works by providing higher GPU utilization.
\tool improves on these works by overlapping inter and intra-node communication operations.
BytePS~\cite{osdi20:byteps} utilizes CPU in heterogenous clusters to improve training, which is complementary to \tool.

\spara{Optimizing Stencil Computations}
Prior works have proposed several DSLs and optimizations for data-parallel stencil computations on CPUs, GPUs, and other accelerators.
Halide~\cite{halide} and Fireiron~\cite{fireiron} separate the algorithm and schedule, which describes the optimizations like fusion, and loop tiling.
TVM~\cite{tvm18} extends Halide for generating optimized compute kernels. 
\textsc{Lift}~\cite{lift-cgo18,lift-cgo17} and PolyMage~\cite{polymage-gpu} automatically optimize stencil computations for a single GPU.
Distributed-Halide~\cite{distributed-halide} extends Halide with scheduling primitives that allow distributing parallelizable dimensions of loops.
\tool extends these works to reason about and compose collective communication with computation, which is crucial for distributed machine learning scenarios.

\spara{Overlapping Computation and Communication}
State-of-the-art works on overlapping \cite{Barigou2017, KOZIRIS20031138,7336201,10.1145/1810085.1810091,10.1007/978-3-319-58667-0_18} use either pipelined execution to overlap communication and computation or non-blocking MPI operations.
Pencil~\cite{sc20:pencil} improves upon these works by performing pipelining within a process and supports computations in multiple connected iteration spaces.
Several techniques distribute tiles and automatically generate communication~\cite{10.1145/2503210.2503289, distributed-halide,8121995}.
Basu et. al.~\cite{6799131} uses overlapped tiling in each process to remove communication between processes.
Denis and Trahay~\cite{7573826} studied the efficiency of overlap.
dCUDA~\cite{dcuda} provides hardware supported overlap.
These works for MPI+OpenMP are valid for CPU based stencil computations that require sends and receives to share the halo regions.
However, unlike \tool, these works do not support overlapping between collectives communication and complex computations like convolutions and matrix multiplications.
\tool supports overlapping multiple computation and communication operations on GPUs without an accelerator.

\section{Conclusion}
\label{sec:conclusion}
This paper introduced \tool, a language to describe distributed machine learning workloads and optimize them across computation and communication boundary. 
We show that \tool{} generated code significantly improves several training and inference times of large language models. 
In the future we plan to automate the optimizations through smart search.


\section{Data Availability Statement}
The artifact for this paper~\cite{coconet-artifact} contains the source code of our implementation of \tool and the benchmarking infrastructure to reproduce all the results in Section~\ref{sec:experiments}.

\begin{acks}
We thank the reviewers and our shepherd, Tyler Sorensen, for their constructive feedback.
This work was partially supported by the National Science Foundation grant CCF-2052696.
\end{acks}

\appendix
\section{Artifact Appendix}

\subsection{Abstract}

This artifact appendix describes how to reproduce results for
standalone experiments in Figure~\ref{fig:bandwidth64GPUs},
~\ref{fig:matmul-overlap}, and ~\ref{fig:pipeline-overlap} and integration 
results in Section~\ref{sec:experiments:bert},~\ref{sec:experiments:model-parallel:integeration}, and ~\ref{sec:experiments:pipeline-parallel:integeration}.
This artifact includes the \tool{} DSL and compiler, and \tool{}'s generated code integrated 
with PyTorch, Megatron-LM, and NVIDIA Bert.
To reproduce the results, the experiments should be executed on a system similar to our experimental system.
However, all experiments can be executed on a system with more than one NVIDIA GPUs.

\subsection{Artifact Check-list (Meta-information)}

{\small
\begin{itemize}
  \item {\bf Program:} \tool{} DSL and compiler written in C++.
  \item {\bf Compilation:} A C++ compiler (\texttt{g++} or \texttt{clang}) to compile \tool{}. A C++ compiler with MPI support (\texttt{mpicxx}) and CUDA compiler (\texttt{nvcc}) to compile generated programs.
  \item {\bf Binary:} Each \tool{} program compiles to a binary that generates an MPI program containing CUDA kernels.
  \item {\bf Data set:} BERT, GPT-2, and GPT-3 training datasets for integration experiments. 
  \item {\bf Run-time environment:} Ubuntu 20.04 with Python 3.7+, CUDA 11.0+, and OpenMPI 4.0+.
  \item {\bf Hardware:} We performed experiments on 16 NVIDIA DGX-2 nodes, i.e., a total of 256 NVIDIA Tesla V100 GPUs. 
  However, the experiments can be executed on any system with two or more GPUs.
  \item {\bf Run-time state:} Python, MPI, and CUDA.
  \item {\bf Execution:} Use \texttt{mpirun} to run the experiments.
  \item {\bf Metrics:} Decrease in execution time of benchmarks.
  \item {\bf Output:} Execution time of each experiment and \tool{} speedup over baselines.
  \item {\bf Experiments:} Execution of standalone experiments and training and inference tasks of BERT, GPT-2, and GPT-3 models.
  \item {\bf How much disk space required (approximately)?:} 100 GB in total. 90\% of the space usage is required for storing dataset.
  \item {\bf How much time will be spent in preparing the workflow (approximately)?:} 1 hour.
  \item {\bf How much time is needed to complete experiments (approximately)?:} 5 hours.
  \item {\bf Publicly available?:} Yes.
\end{itemize}
}

\subsection{Description}

\subsubsection{How to Access}

The \tool{} implementation and the benchmarking infrastructure used in our evaluation are publicly available as the artifact~\cite{coconet-artifact}. 
This artifact contains a zip file with two directories: (i) \texttt{coconet}, which is the implementation of \tool{}, and (ii) \texttt{coconet-experiments}, which is the benchmarking infrastructure.
Latest versions of these directories are available at \url{https://github.com/parasailteam/coconet} and \url{https://github.com/parasailteam/coconet-experiments}.

\subsubsection{Hardware Dependencies}
All benchmarks can be executed on a distributed system with two or more NVIDIA GPUs.
However, our results will be reproducible on the evaluation system described in Section~\ref{sec:experiments}.

\subsubsection{Software Dependencies}
Our experiments require a system running Ubuntu 20.04 with Python 3.8+ and CUDA 11.0+.
Prerequisites and their installation procedure is described in \texttt{README.md} files of \texttt{coconet} and \texttt{coconet-experiments} directories.

\subsubsection{Data Sets}
The standalone benchmarks (Figure~\ref{fig:bandwidth64GPUs},~\ref{fig:matmul-overlap}, and ~\ref{fig:pipeline-overlap}) do not require any dataset.
Datasets required for executing experiments in Section~\ref{sec:experiments:bert},~\ref{sec:experiments:model-parallel:integeration}, and ~\ref{sec:experiments:pipeline-parallel:integeration} can be obtained by following \textit{Dataset} section of \texttt{README.md} in \texttt{coconet-experiments}.
\subsection{Installation}
Following instructions have been tested with Ubuntu 20.04.

\paragraph{Standalone Experiments Dependencies} Install dependencies by following the \textit{Prerequisites} section in \texttt{README.md} file of \texttt{coconet} directory.

\paragraph{Integration Experiments Dependencies} Follow the \textit{Prerequisites} section in \texttt{README.md} file of \texttt{coconet-experiments} directory to build PyTorch and install all dependencies for Megatron-LM and NVIDIA Bert.

\subsection{Experiment Workflow}
\subsubsection{Standalone Experiments}
\label{appendix:sec-standalone}
This section describe how to execute standalone experiments of Section~\ref{sec:experiments} and produce results for Figure~\ref{fig:bandwidth64GPUs}, Figure~\ref{fig:matmul-overlap}, and Figure~\ref{fig:pipeline-overlap}.
All of these experiments will take 1 hour combined.
\begin{enumerate}
  \item Install all \tool prerequisites in \texttt{coconet/README.md}.
  \item The \texttt{experiments/} directory contains all scripts for standalone experiments.

{\footnotesize
\begin{lstlisting}[language=bash]
$ cd coconet/experiments/
\end{lstlisting}
}

  \item Since all our experiments uses MPI to run the executable on all GPUs,  set the environment variable \texttt{NPROC} to the number of GPUs in the system. In our experiments, we set \texttt{NPROC} to 256 as follows:
  
  {\footnotesize
\begin{lstlisting}[language=bash]
$ export NPROC=256
\end{lstlisting}
}
\textbf{Note}: Setting \texttt{NPROC} to a value more than the number of GPUs in a system can lead to failed experiments.

  \item If the experiments are performed on a system with multiple nodes then additional arguments to \texttt{mpirun} can be passed by setting the \texttt{MPI\_ARGS} environment variable.
\end{enumerate}

\paragraph{Data-Parallel Experiments}
\begin{enumerate}
  \item To execute standalone data parallel experiments execute \texttt{data-parallel-exp.py}. This script takes a directory to store the results as an argument.
  Additionally, the script requires \texttt{MASTER\_ADDR} and \texttt{MASTER\_PORT} to be passed as \texttt{MPI\_RUN\_ARGS}. If the experiments are done on a single system, then it is common to set \texttt{MASTER\_ADDR=127.0.0.1} and \texttt{MASTER\_PORT=10000}.

  {\footnotesize
\begin{lstlisting}[language=bash]
$ export MPI_ARGS="-x MASTER_ADDR=127.0.0.1"
$ export MPI_ARGS="$MPI_ARGS -x MASTER_PORT=10000"
$ python data-parallel-exp.py results/
\end{lstlisting}
}

  The above execution of script will execute all data parallel executables and store the results in the \texttt{results} directory.

  \item Generate both graphs of Figure~\ref{fig:bandwidth64GPUs} by executing the script \texttt{gen-data-parallel-graphs.py}. This script takes the directory with results generated in the previous step as an argument.
  
{\footnotesize
\begin{lstlisting}[language=bash]
$ python gen-data-parallel-graphs.py results/
\end{lstlisting}
}

Graphs are stored in two files of \texttt{experiments} directory: \texttt{results-adam-fp16.pdf} and \texttt{results-lamb-fp16.pdf}.
\end{enumerate}

\paragraph{Model-Parallel Experiments}
\begin{enumerate}
  \item To execute standalone model-parallel experiments execute \texttt{model-parallel-exp.py}. Similar to the previous script, this script also takes a directory to store results as its argument.
  
  {\footnotesize
\begin{lstlisting}[language=bash]
$ python model-parallel-exp.py results/
\end{lstlisting}
}

  The script will execute all model parallel executables and stores the results in the \texttt{results} directory.
  \item Generate Figure~\ref{fig:matmul-overlap} by executing following script. This script will take above results directory as its argument. 
  
{\footnotesize
\begin{lstlisting}[language=bash]
$ python gen-model-parallel-graphs.py results/
\end{lstlisting}
}

Graph is stored as \texttt{results-model-parallel.pdf}.
\end{enumerate}

\paragraph{Pipeline-Parallel Experiments}
\begin{enumerate}
  \item To execute standalone pipeline-parallel experiments execute \texttt{pipeline-parallel-exp.py}. This script also requires a directory to store results as its command line argument.

  {\footnotesize
\begin{lstlisting}[language=bash]
$ python pipeline-parallel-exp.py results/
\end{lstlisting}
}

  Above execution of the script will execute all pipeline parallel executables and store the results in \texttt{results} directory.
  \item To generate Figure~\ref{fig:pipeline-overlap} execute the script \\ \texttt{gen-pipeline-parallel-graphs.py}. This script takes the directory containing above results as its argument.
  
{\footnotesize
\begin{lstlisting}[language=bash]
$ python gen-pipeline-parallel-graphs.py results/
\end{lstlisting}
}

The graph is stored in \texttt{results-model-parallel.pdf}.
\end{enumerate}

\subsubsection{Integration Experiments}
\label{appendix:sec-integration}
In this section, we will execute the integration experiments of Section~\ref{sec:experiments:bert},~\ref{sec:experiments:model-parallel:integeration}, and~\ref{sec:experiments:pipeline-parallel:integeration}.

\paragraph{Prerequisites} Install prerequisites and obtain dataset by following the steps in \texttt{coconet-experiments/README.md}.

\paragraph{Data-Parallel Training}
Go to \texttt{Nvidia-Bert} directory and execute \texttt{coconet-experiments.py}.

{\footnotesize
\begin{lstlisting}[language=bash]
$ cd NV-BERT 
$ python coconet-experiments.py
\end{lstlisting}
}

This script will execute data parallel training experiments and then print Table~\ref{tab:bert-results}.
This experiment will take 1 hour to complete.
This script contains maximum batch sizes supported by each implementation for our evaluation system of 256 Tesla V100 GPUs.
It is possible that for a different system the maximum batch size will be different.
The batch size dictionary in \texttt{coconet-experiments.py} can be modified to find maximum batch size for underlying system.

\paragraph{Model-Parallel Inference} 
Go to \texttt{MegatronLM-Model-Parallel} directory and execute \texttt{coconet-experiments.py}.
{\footnotesize
\begin{lstlisting}[language=bash]
$ cd MegatronLM-Model-Parallel
$ python coconet-experiments.py
\end{lstlisting}
}
This script will execute model parallel inference experiments and then print the values in Section~\ref{sec:experiments:model-parallel:integeration}.
This experiment will take less than 30 minutes to complete.

\paragraph{Pipeline-Parallel Inference} 
Execute \texttt{coconet-experiments.py} in the directory
\texttt{MegatronLM-Pipeline-Parallel}.
{\footnotesize
\begin{lstlisting}[language=bash]
$ cd MegatronLM-Pipeline-Parallel
$ python coconet-experiments.py
\end{lstlisting}
}
This script will execute pipeline parallel inference experiments and then print the table in Section~\ref{sec:experiments:pipeline-parallel:integeration}.
This experiment will take 3 hour to complete.
\subsection{Evaluation and Expected Results}

\paragraph{Standalone Experiments} The figures generated by the experiments of Section~\ref{appendix:sec-standalone} can be matched with the figures: \ref{fig:bandwidth64GPUs}, \ref{fig:matmul-overlap}, and \ref{fig:pipeline-overlap}.

\paragraph{Integration Experiments}
The results generated in experiments of Section~\ref{appendix:sec-integration} can be matched with the results in Section~\ref{sec:experiments:bert},~\ref{sec:experiments:model-parallel:integeration}, and~\ref{sec:experiments:pipeline-parallel:integeration}.




\balance
\bibliographystyle{ACM-Reference-Format}
\interlinepenalty=10000
\bibliography{paper}

\end{document}